\documentclass[iop]{emulateapj}
\usepackage{graphicx}
\usepackage{color}
\usepackage{amssymb}
\usepackage{amsmath}
\usepackage{epstopdf}
\usepackage{xspace}
\newcommand{\fref}{Figure~\ref}
\newcommand{\tref}{Table~\ref}
\newcommand{\sref}{Section~\ref}
\newcommand{\rg}{\ensuremath{R_{\rm g}}}
\newcommand{\xmm}{{\em XMM--Newton}\xspace}
\newcommand{\nus}{{\em NuSTAR}\xspace}
\newcommand{\suz}{{\em Suzaku}\xspace}
\newcommand{\cha}{{\em Chandra}\xspace}
\newcommand{\msun}{\ensuremath{M_{\odot}}\xspace}
\newcommand{\mdot}{\ensuremath{\dot{M}}\xspace}
\def\arcsec{\ensuremath{\,^{\prime\prime}}\xspace}
\def\arcmin{\ensuremath{\,^{\prime}}\xspace}
\def\degree{\ensuremath{\,^{\rm o}}\xspace}
\newcommand{\dkbbfth}{{\tt dkbbfth}\xspace}
\newcommand{\diskbb}{{\tt diskbb}\xspace}
\newcommand{\diskpbb}{{\tt diskpbb}\xspace}
\newcommand{\comptt}{{\tt comptt}\xspace}
\newcommand{\optxagnf}{{\tt optxagnf}\xspace}
\newcommand{\slimdisk}{{\tt slimdisk}\xspace}
\newcommand{\tbnew}{{\tt tbnew}\xspace}
\newcommand{\tbnewfeo}{{\tt tbnew\_feo}\xspace}
\newcommand{\tbabs}{{\tt tbabs}\xspace}

\newcommand{\berk}{4}
\newcommand{\dtu}{5}
\newcommand{\lawr}{6}
\newcommand{\camb}{7}
\newcommand{\colu}{8}
\newcommand{\gsfc}{9}
\newcommand{\mich}{10}
\newcommand{\jpl}{11}

\shorttitle{NGC~1313 X-1 and X-2 with NuSTAR and XMM-Newton}
\shortauthors{Bachetti et al.}

\begin{document}
\title{The Ultraluminous X-ray sources NGC~1313 X-1 and X-2:\\a broadband study with \nus and \xmm }
\author{Matteo Bachetti\altaffilmark{1,2}}\email{matteo.bachetti@irap.omp.eu}
\author{Vikram Rana\altaffilmark{3}}
\author{Dominic J. Walton\altaffilmark{3}}
\author{Didier Barret\altaffilmark{1,2}}
\author{Fiona A. Harrison\altaffilmark{3}}
\author{Steven E. Boggs\altaffilmark{\berk}}
\author{Finn E. Christensen\altaffilmark{\dtu}}
\author{William W. Craig\altaffilmark{\berk,\lawr}}
\author{Andrew C. Fabian\altaffilmark{\camb}}
\author{Felix F\"urst\altaffilmark{3}}
\author{Brian W. Grefenstette\altaffilmark{3}}
\author{Charles J. Hailey\altaffilmark{\colu}}
\author{Ann Hornschemeier\altaffilmark{\gsfc}}
\author{Kristin K. Madsen\altaffilmark{3}}
\author{Jon M. Miller\altaffilmark{\mich}}
\author{Andrew F. Ptak\altaffilmark{\gsfc}}
\author{Daniel Stern\altaffilmark{\jpl}}
\author{Natalie A. Webb\altaffilmark{1,2}}
\author{William W. Zhang\altaffilmark{\gsfc}}

\altaffiltext{1}{Universit\'e de Toulouse; UPS-OMP; IRAP; Toulouse, France}
\altaffiltext{2}{CNRS; Institut de Recherche en Astrophysique et Plan\'etologie; 9 Av. colonel Roche, BP 44346, F-31028 Toulouse cedex 4, France}
\altaffiltext{3}{Cahill Center for Astronomy and Astrophysics, Caltech, Pasadena, CA 91125}
\altaffiltext{\berk}{Space Sciences Laboratory, University of California, Berkeley, CA 94720, USA}
\altaffiltext{\dtu}{DTU Space, National Space Institute, Technical University of Denmark, Elektrovej 327, DK-2800 Lyngby, Denmark}
\altaffiltext{\lawr}{Lawrence Livermore National Laboratory, Livermore, CA 94550, USA}
\altaffiltext{\camb}{Institute of Astronomy, University of Cambridge, Madingley Road, Cambridge CB3 0HA, UK}
\altaffiltext{\colu}{Columbia Astrophysics Laboratory, Columbia University, New York, NY 10027, USA}
\altaffiltext{\gsfc}{NASA Goddard Space Flight Center, Greenbelt, MD 20771, USA}
\altaffiltext{\mich}{Department of Astronomy, University of Michigan, 500 Church Street, Ann Arbor, MI 48109-1042, USA}
\altaffiltext{\jpl}{Jet Propulsion Laboratory, California Institute of Technology, Pasadena, CA 91109, USA}

\begin{abstract}
We present the results of \nus and \xmm observations of the two ultraluminous X-ray sources NGC~1313 X-1 and X-2. The combined spectral bandpass of the two satellites enables us to produce the first spectrum of X-1 between 0.3 and 30\,keV, while X-2 is not significantly detected by \nus above 10\,keV. The \nus data demonstrate that X-1 has a clear cutoff above 10\,keV, whose presence was only marginally detectable with previous X-ray observations. This cutoff rules out the interpretation of X-1 as a black hole in a standard low/hard state, and it is deeper than predicted for the downturn of a broadened iron line in a reflection-dominated regime. The cutoff differs from the prediction of a single-temperature Comptonization model. Further, a cold disk-like black body component at $\sim0.3$\, keV is  required by the data, confirming previous measurements by \xmm only. We observe a spectral transition in X-2, from a state with high luminosity and strong variability to a lower-luminosity state with no detectable variability, and we link this behavior to a transition from a super-Eddington to a sub-Eddington regime.
\end{abstract}


\maketitle

%
%

\section{Introduction}

Ultraluminous X-ray sources (ULXs) are off-nuclear point-like sources with apparent X-ray luminosities exceeding the Eddington limit for stellar-mass black holes (StMBHs). Their high luminosity can be due to yet-unknown mechanisms of super-Eddington accretion on a StMBH (or beamed emission from it), or the presence of a black hole (BH) with a high mass, such as an intermediate-mass black hole (IMBH). While for luminosities $>10^{41} {\rm erg\,s^{-1}}$ the identification with an IMBH is most probable, as was shown for the source HLX-1 \citep{Farrell+09}, for lower luminosities both mechanisms can apply. Convincing evidence for super-Eddington accretion has been reported for two ULXs in M31 \citep{Middleton+12,Middleton+13}. See \citet{Roberts07} and \citet{FengSoria} for reviews.

ULX spectra below 10\,keV have been thoroughly investigated  \citep[see, e.g.,][]{Gladstone+09} with \xmm \citep{xmm01}, \suz \citep{suzaku07} and \cha \citep{chandra02}. Their X-ray spectral shape does not match that of known BHs, in the mass range from 10 to millions of solar masses (see \citealt{Done+07} for a review of standard BHs). A spectral break below 10\,keV has been observed in most ULXs \citep{Stobbart+06,Gladstone+11}, together with a disk-like black body component at low temperatures ($\lesssim 0.3$\,keV). This latter component, if produced by a standard disk reaching the proximity of the BH, would imply masses above $\sim100\,\msun$, and thus the presence of an IMBH \citep{Miller+03,Miller+04}. But the temperature-luminosity relation for this component does not match that expected in standard accretion disks in the soft state, where the disks extend to the innermost stable circular orbit (see, e.g., \citealt{KajavaPoutanen09}, \citealt{FengSoria} for a review). This relation can be partially recovered in some cases by assuming a constant absorption column between the observations \citep{Miller+13} or using non-standard disk models \citep{Vierdayanti+06}. Also, the cutoff is at much lower temperature than is expected in standard BH hard states \citep{Done+07}. Some authors associate the low-temperature disk-like component with the presence of an optically thick corona that blocks the inner part of the disk, so that the visible part of the disk has a much lower temperature \citep{Gladstone+09}. Others suggest that it might come from a strong outflow \citep[e.g.][]{King+04} or be the result of blurred line emission from highly ionized, fast-moving gas \citep{GoncalvesSoria06}. 

From X-ray data below 10\,keV it is impossible to distinguish between a cutoff and a downturn produced by the imperfect fit of the continuum due to the presence of a broadened iron complex in a reflection-dominated regime \citep{CaballeroGarcia+10,Gladstone+11}. The difference becomes clear above 10\,keV \citep[see, e.g.,][]{Walton+11}, in a region of the spectrum that past sensitive satellites could not explore.


The {\em Nuclear Spectroscopic Telescope Array} \citep[\nus;][]{nustar13}, launched in 2012 June, with its focusing capabilities, large bandpass between 3 and 80\,keV and large effective area, represents the ideal complement to \xmm (given the similar effective area between 5 and 10\, keV and spectral capabilities). 
X-rays are focused by multilayer-coated grazing incidence optics onto two independent focal plane modules, called Focal Plane Module A and B (hear after FPMA and FPMB). Each focal plane contains four cadmium zinc telluride detectors. The spatial resolution is 58\arcsec half-power diameter and 18\arcsec FWHM. NuSTAR is therefore a powerful tool for studying ULX broad band X-ray spectra. 
Since the launch of the satellite, we have observed a sample of luminous ($L_{\rm x}\sim10^{40}\,{\rm erg\,s^{-1}}$), close-by ($d\lesssim{10}\,{\rm Mpc}$) and hard (showing X-ray power law photon index $\Gamma \lesssim 2$ below 10\,keV) ULXs simultaneously with \nus and \suz or \xmm, producing the first ULX spectra extending over the range 0.3  and  30\,keV.

In this paper, we describe the results obtained for the two ULXs in the spiral galaxy NGC~1313, ($d\sim 4.13$\,Mpc, \citealt{Mendez+02}). These two ULXs are among the brightest, hardest and closest ULXs \citep[][]{Swartz+04,Walton+11cat}, and therefore they are ideal targets for our program. They are known to show spectral variability below 10\,keV \citep{Feng+06,Dewangan+10,Pintore+12}. Significant variability at high fluxes has also been observed in both sources \citep{Heil+09}. 

In \sref{sec:obs} we  describe  the observations done, in \sref{sec:data} we provide some details on data reduction, then in \sref{sec:spectral} and \sref{sec:timing} we discuss the spectral and timing analysis of the two sources, and finally we discuss the results.

%
%

\section{The observations}\label{sec:obs}
\begin{deluxetable}{l c c c }
\tablecolumns{4}
\tablewidth{\columnwidth}
\tabletypesize{\scriptsize}
\tablecaption{Summary of the Data used in This Paper\label{tab:sum}}
\tablehead{\colhead{Camera} & \colhead{Exposure (ks)} & \colhead{X-1 Counts} & \colhead{X-2 Counts} }
\tablecomments{Values in parentheses are background counts, scaled to the source region size.\\
$^1$X-1 on detector edge; $^2$X-2 on detector edge}
\startdata
\cutinhead{Epoch 1 -- 2012 Dec 16}
FPMA      & 100.9       & 3314 (386.2) & 2336 (1074.2)   \\
FPMB      & 100.8       &  3444 (473.0) & 2504 (1076.3)   \\
EPIC-pn   & 93.8       & 74002 (1523.4) & 52603 (1029.1)  \\
EPIC-MOS1$^1$ & 114.5       & 27785 (309.2) & 21233 (318.8) \\
EPIC-MOS2 & 115.1       & 29917 (339.5) & 21304 (323.6)  \\
\cutinhead{Epoch 2 -- 2012 Dec 21--22}
FPMA      & 127.0       & 4166 (472.9) & 1898 (1333.3)   \\
FPMB      & 127.0       & 4237 (584.2) & 1918 (1517.4)  \\
EPIC-pn$^2$   & 79.2        & 70605 (881.6) & 22925 (759.2)\\
EPIC-MOS1$^1$ & 116.0       & 20439 (411.4) & 12378 (384.8)\\
EPIC-MOS2 & 121.8       & 32796 (439.6) & 13593 (326.8) 
\enddata
\end{deluxetable}

During this campaign, we observed NGC~1313 with \xmm and \nus two times, as summarized in \tref{tab:sum}. Observations were executed with a separation of about a week, to search for variability. The two ULXs are separated by about 7\arcmin and can be observed simultaneously by \xmm and \nus. 
We chose to place X-1 close to the optical axis. It was not possible to keep both ULXs close to the optical axis of \nus, so we chose to obtain the best spectral quality for at least one of them rather than reducing the quality for both. X-1 is historically brighter and harder than X-2 \citep{Pintore+11}, and we estimated that the addition of \nus data would yield more valuable new information for this source.

%
%

\section{Data Reduction}\label{sec:data}
\subsection{\nus Data}\label{sec:nusproc}\label{sec:contam}

\nus data were processed using the version 1.0.1 of the \nus data analysis system, ({\em NuSTAR DAS}). 
The {\em NuSTAR DAS} tools are divided in two main parts: the preprocessing pipeline ({\tt nupipeline}) that produces the L1 filtered files, and the products pipeline ({\tt nuproducts}) that is used to extract spectra, lightcurves and other high-level products. 

We ran {\tt nupipeline} on all observations with the default options for good time interval filtering, and produced cleaned event files. We then ran {\tt nuproducts} using a 30\arcsec extraction region around X-1 (see \sref{sec:contam} for the details) and a 60\arcsec extraction region around X-2, and a for background an 80\arcsec extraction region in the same detector as the source, further than 1' away to avoid contributions from the point-spread function (PSF) wings. We applied standard PSF, alignment and vignetting corrections. Spectra were rebinned in order to have at least 20 counts bin$^{-1}$ to ensure the applicability of the $\chi^2$ statistics, and in some cases to 50 counts bin$^{-1}$ in order to reduce computation times in particularly complicated models. 

As it turned out, \nus data of X-2 produced very poor spectral information above $\sim 10$\,keV. Besides being very faint, the \nus data were likely to be affected by response degradation due to the off-axis position of the source, and a very uncertain background level due to the \nus sloping aperture background. We decided not to use them for the next steps of the analysis.

\begin{figure*}
\centering
$$
\begin{array}{ccc}
\includegraphics[width=3in]{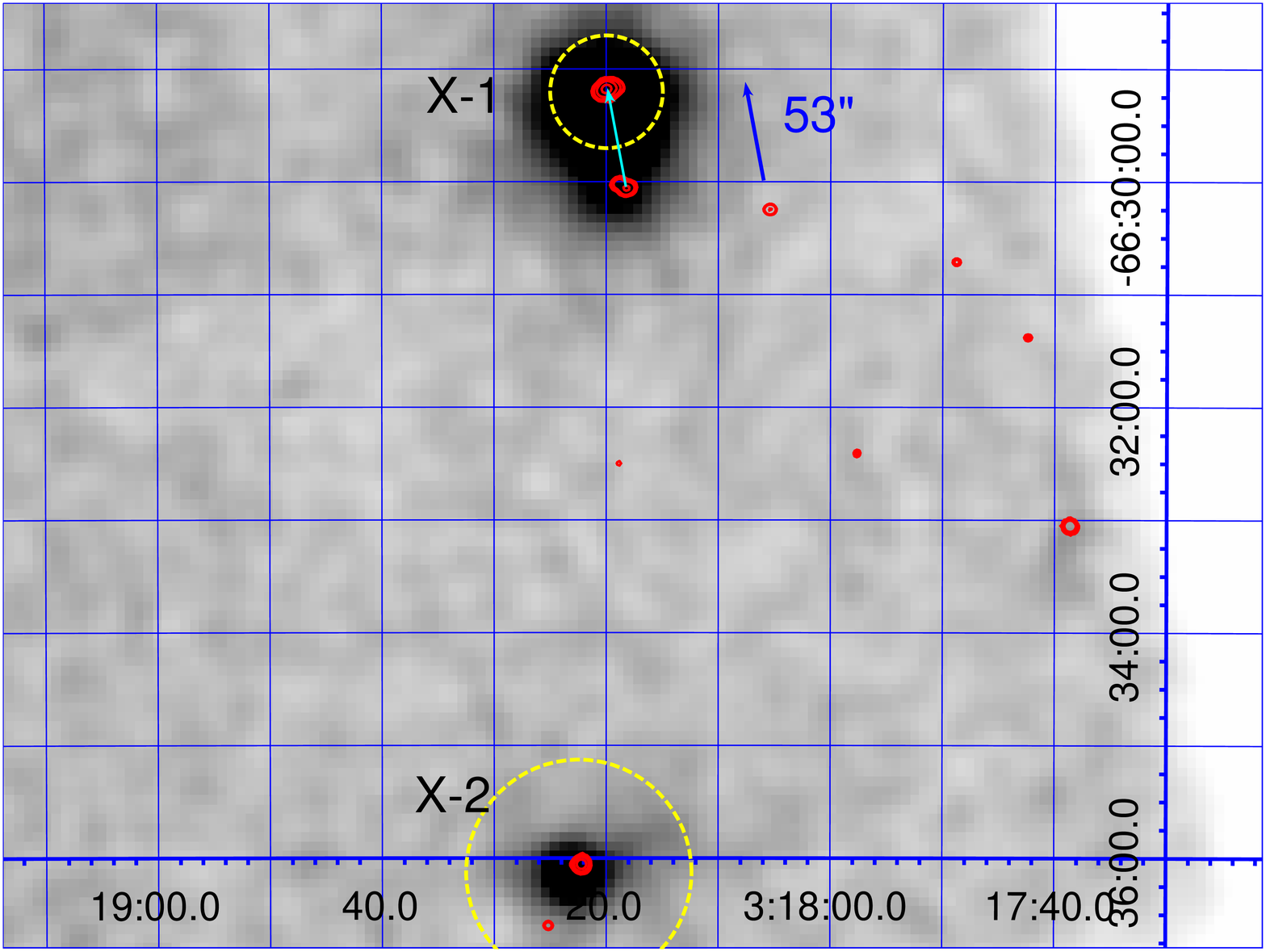} &
\includegraphics[width=3in]{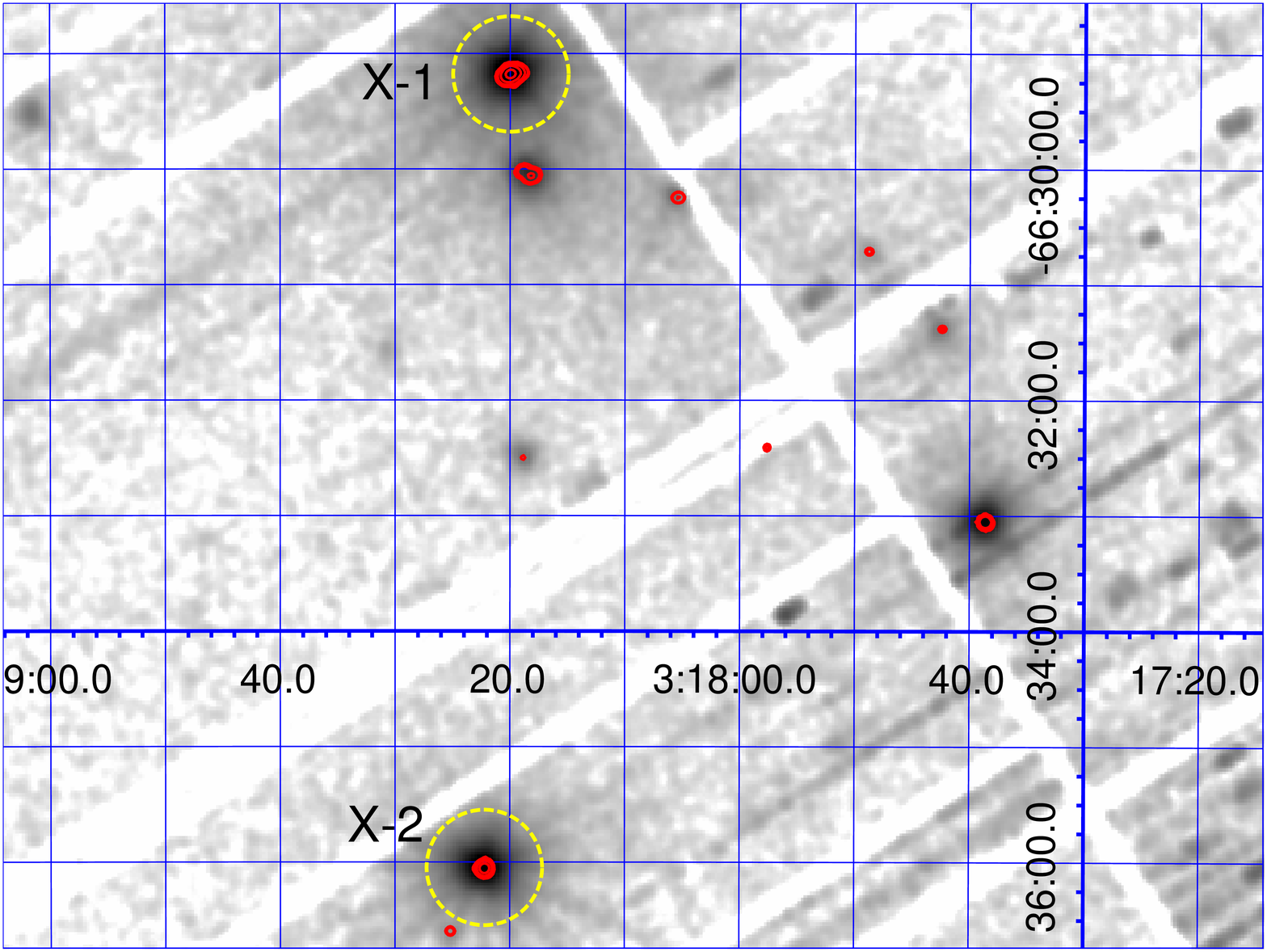} &
\end{array}
$$
\caption{(Left) \nus and (right) EPIC-pn images of the two ULXs, produced with DS9 \citep{ds9}. Data are from the whole energy bands of the detectors. \cha contours corresponding to the ULXs and possible contaminants are shown in red. Yellow dashed regions are the extraction regions used for analysis. The radius of the region around X-1 is 30\arcsec in both cases in order to avoid the contaminating source about 50\arcsec SE of the source. For X-2, instead, it is 60\arcsec in \nus and 30\arcsec for \xmm.}
\label{fig:contours}
\end{figure*}

\begin{figure}
\includegraphics[angle=270,width=\columnwidth]{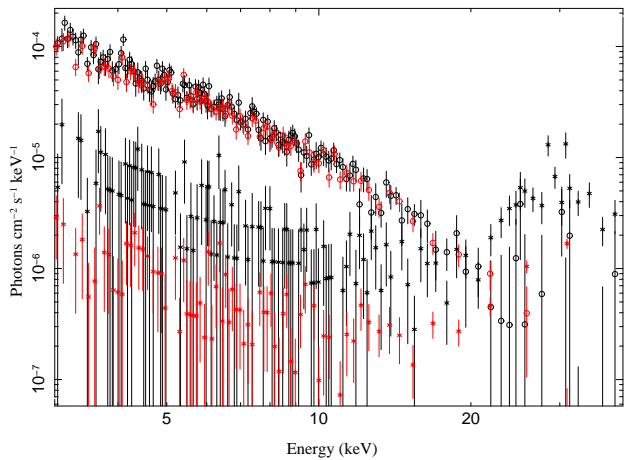}
\caption{\nus spectrum of X-1, rebinned to 30 counts bin$^{-1}$; black corresponds to an $80\arcsec$ extraction region, red to $30\arcsec$ (which excludes the contaminating region 53\arcsec SE of X-1). Circles label the source spectra, while ``x''s mark the background spectra. There is no significant change in the spectrum between 10 and 30\,keV, but the larger extraction region is much more affected by background. Also, below 10\,keV there is some very small deviation.}
\label{fig:contam}
\end{figure}

As can be seen in \fref{fig:contours}, NGC~1313 X-1 has a nearby contaminating source separated by $\sim 53\arcsec$ that is not clearly resolved by \nus. While the source is outside the \xmm PSF of X-1 and it is quite easy to avoid it through the choice of a small extraction region, the evaluation of its possible effect on \nus data is less straightforward, due to the larger PSF. The contaminating source has a flux $\sim10$ times lower than X-1 in the \xmm band, but the \nus PSF of X-1 appears elongated towards the contaminating source. To evaluate the effects of this source, we produced \nus spectra with two different extraction regions, one including the nearby source (radius 80\arcsec) and one not including it (radius 30\arcsec). As shown in \fref{fig:contam}, the two spectra do not differ substantially between 10\,keV and the intersection of source and background levels, while there is some minor deviation at lower energy, in the \xmm band. The best-fit power laws below 10\,keV in the two datasets are marginally compatible (spectral index  $2.08\pm0.08$ in the first and $1.9\pm0.1$ in the second; quoted errors are 90\% confidence limits). We chose to use the smaller extraction region for precaution. This analysis shows that the residual effect is negligible if the 30\arcsec extraction region is used. In the following analysis, we only consider spectra below 30\,keV where the source is stronger than, or compatible with, the background.

\subsection{\xmm Data}

The \xmm data reduction was carried out with the \xmm Science Analysis System (SAS v12.0.1). 
We produced calibrated event files with {\tt epproc} and {\tt emproc}, created custom good time interval files to filter out periods of high background according to the prescription in the SAS manual, and selected only {\tt \#XMMEA\_EP \&\& PATTERN<4} events for EPIC-pn and {\tt \#XMMEA\_EM \&\& PATTERN<12} events for EPIC-MOS cameras. We also filtered the events along detector gaps through {\tt FLAG==0}.
The resulting event files were then filtered with a 30\arcsec region around the two ULXs. Background events were selected in each detector in regions with no detector edges, bad pixels or visible sources.

Spectra were extracted for all three cameras, unless the source was in a detector gap (see \tref{tab:sum}). We used {\tt fselect} for spectral extraction, and ancillary responses and redistribution matrices were created with {\tt arfgen}  and {\tt rmfgen}, with the new ELLBETA PSF correction enabled.
Spectra were finally rebinned with {\tt grppha} in order to have at least 20 counts bin$^{-1}$.

%
%

\section{Spectral analysis}\label{sec:spectral}
\subsection{Software Tools and General Procedure}
Spectral analysis was carried out with the Interactive Spectral Analysis System \citep[ISIS;][]{Houck+00}. We chose this software over the more commonly used X-ray spectral fitting package XSPEC \citep{Arnaud96} because of its scriptability and the transparent use (in multicore computers) of parallel processing during confidence region calculation and parameter space searching, while being able to use all XSPEC models, table and local models\footnote{In the following sections we will show several {\em unfolded} spectra (i.e. spectra corrected for the response and thus ideally equal to the ``real'' spectrum of the source). In ISIS, the calculation of unfolded spectra is done in a model-independent way by using the response matrices, as opposed to XSPEC where the calculation of these spectra is performed through the distance of data points from the model. This calculation is less statistically robust, and is used only for display purposes.  Model fitting and residuals are calculated in the usual way, by applying the response matrix to the model and comparing to the uncorrected detector counts. See more details in \cite{Nowakrant}.}.

To model neutral absorption we used the \tbnew model\footnote{\url{http://pulsar.sternwarte.uni-erlangen.de/wilms/research/tbabs/}}, the new version of \tbabs \citep{Wilms+00} featuring higher spectral resolution and, also importantly, much faster computation due to caching techniques (see linked Web site for the details). This model can be used in different ways, by including custom abundances of a large number of elements. We use the simplest version, \tbnewfeo, including only the abundances of iron and oxygen besides the usual hydrogen column $n_{\rm H}$, and fixing the abundances of all elements to the standard values from \citet{Wilms+00}. We use the cross sections from \citet{Verner+96}. The hydrogen column we measure from our fits is at least $\sim5$ times higher than the Galactic values taken from \citet{Kalberla+05}. Therefore for simplicity we use only one component for modeling absorption instead of the two that would be necessary were the values comparable.

When jointly fitting \nus and \xmm data, we first fit a {\tt constant*cutoffpl} model between 5 and 10 keV, with the constant for EPIC detectors fixed to 1 and the others left free\footnote{The cross-calibration between pn and MOS\{1,2\} is negligible with respect to the one between pn and FPM in our data.}, to determine a cross-calibration constant that we fix for the subsequent fits. This takes into account residual cross-calibration between \xmm and \nus and the possible mismatches due to non-strictly simultaneous observations.

\subsection{NGC~1313 X-1}\label{sec:specx1}
\begin{figure}
\centering
\includegraphics[origin=c,angle=270,width=\columnwidth]{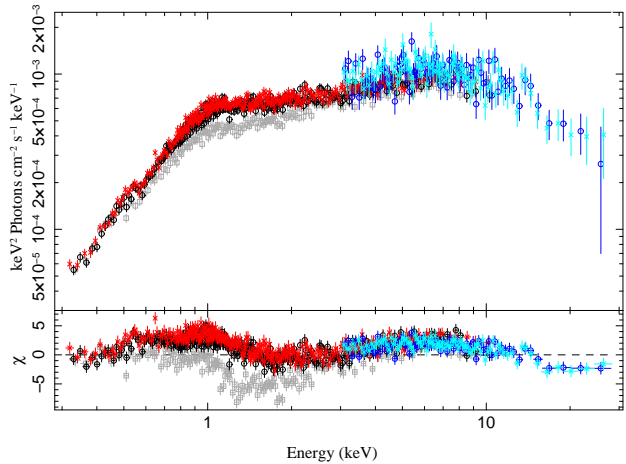}
\caption{EPIC-pn and \nus unfolded spectrum of NGC~1313 X-1 during the two observations, and residuals with respect to the best-fit absorbed power law in the \xmm band. Black and red points are EPIC-pn data, and blue and cyan FPMA data.  Circles indicate the first observation, crosses indicate the second. The spectrum shows a soft excess and a cutoff, as observed in this source when in its low-flux state. The archival 2006 October \xmm observation is also plotted (grey squares) for comparison. Data are rebinned to 200 counts bin$^{-1}$ (EPIC-pn) and 30 counts bin$^{-1}$ (FPMA) for visual purposes. The spectral shape does not change significantly between the two observations and is qualitatively similar to the archival spectrum, with a slightly higher flux. The slight misalignment between \nus and \xmm data is due to residual cross-calibration and possibly the non perfect simultaneity of the observations.}
\label{fig:x1overview}
\end{figure}

\begin{figure*}
\centering
\includegraphics[width=\columnwidth]{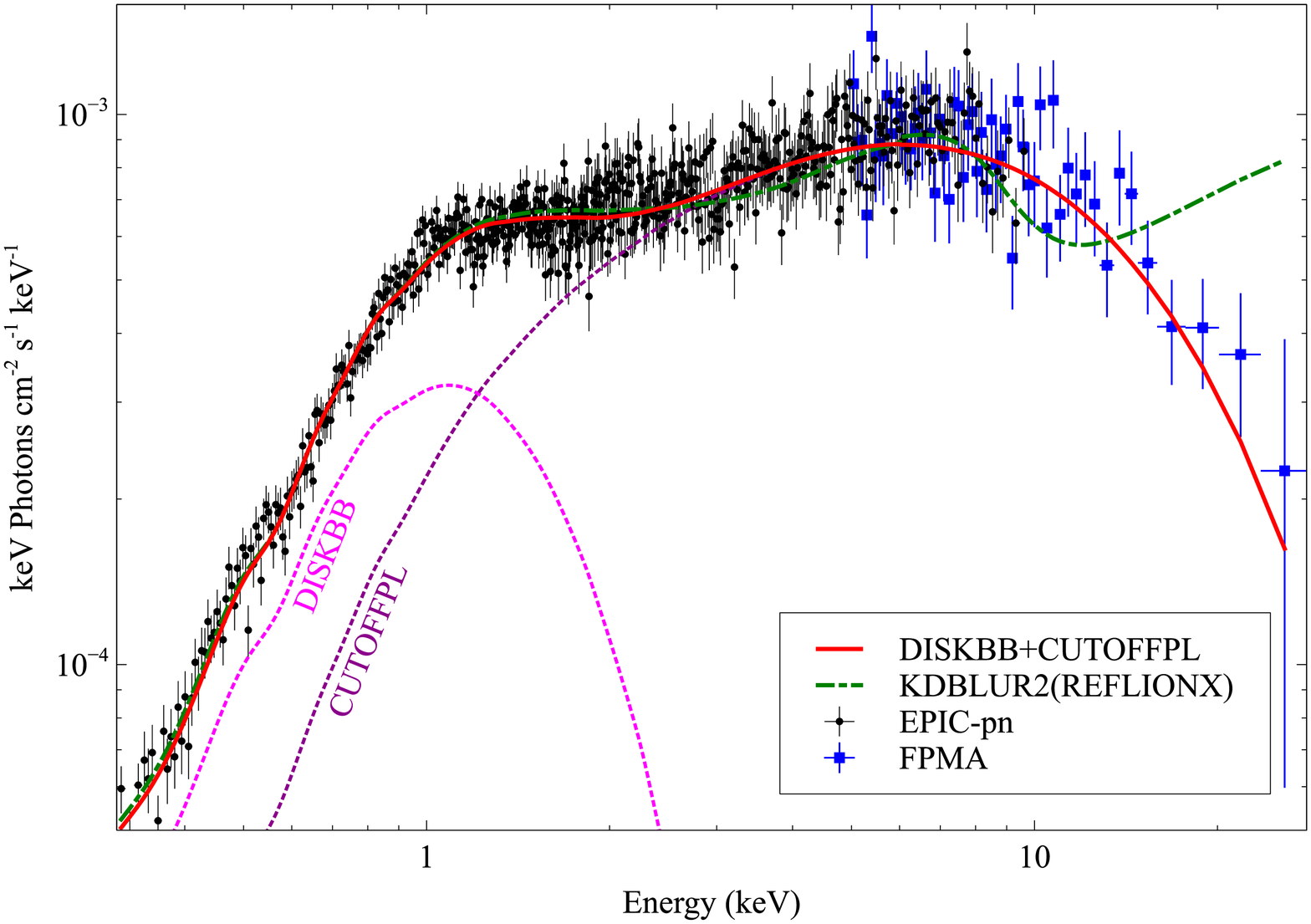}
\includegraphics[width=\columnwidth]{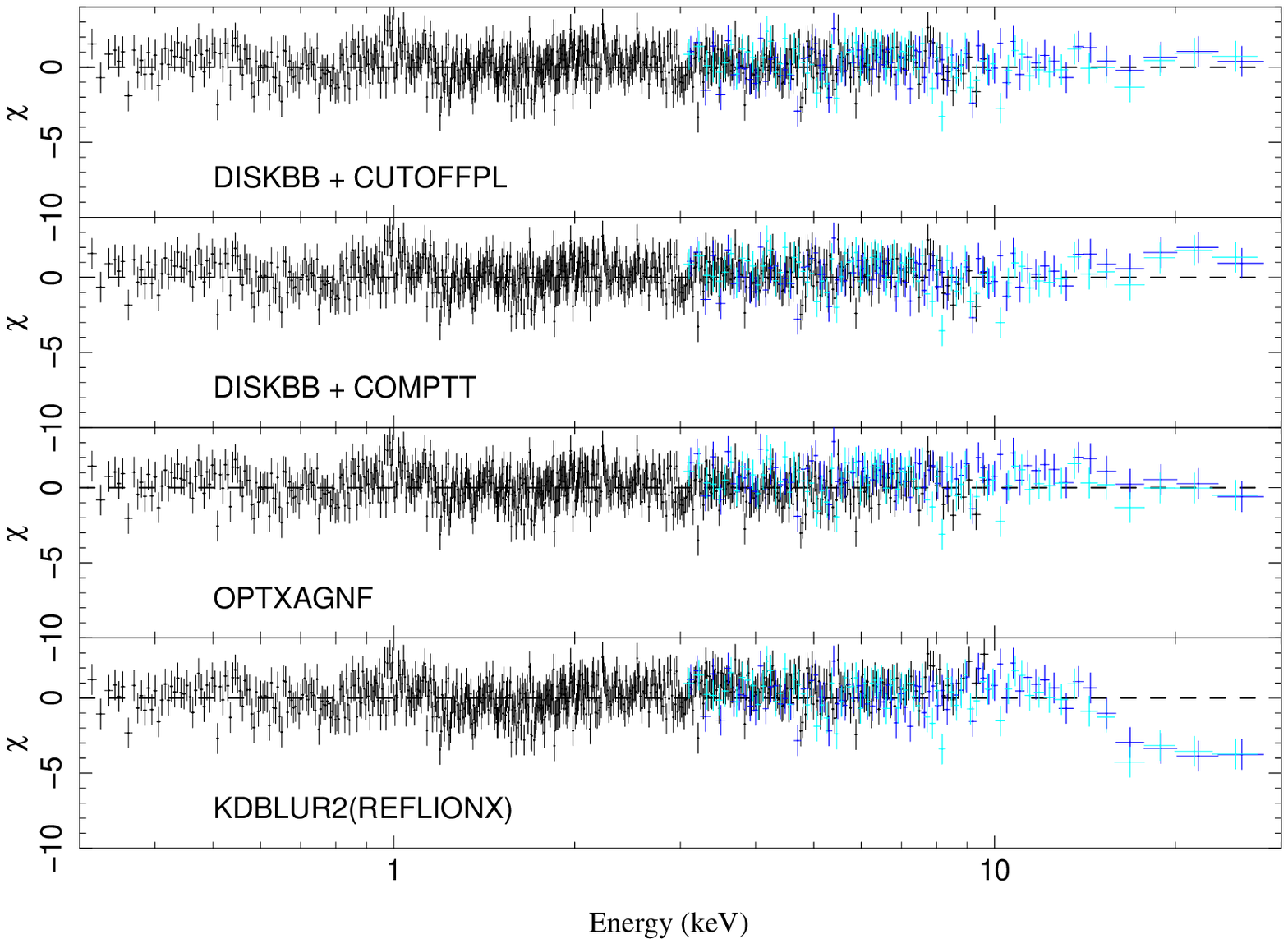}
\caption{({Left}) The unfolded spectrum of the second observation of X-1. Black circles represent EPIC--pn data, and blue squares represent FPMA data normalized with the cross-normalization constant. We superimpose the best-fit {\tt diskbb+cutoffpl} model (red, solid), its single components (dashed), and the best-fit reflection model (green, dash-dotted). ({Right}) residuals from a selection of models listed in Tables~\ref{tab:fitx1} and~\ref{tab:fitoptx1}. Black is EPIC--pn, blue FPMA and light blue FPMB. The models are calculated with all available detectors, but for clarity only pn data are plotted for \xmm. Also for clarity, data have been rebinned to 30 counts bin$^{-1}$. The red model overplotted to the data in the spectrum shows the best fit with a reflection-dominated spectrum. Note that this model works very well in the \xmm band, and the downturn produced by the iron line around 10\,keV is able to fit the cutoff below 10\,keV, but the addition of \nus data clearly rules it out.}
\label{fig:x1multi}
\end{figure*}

\begin{figure}
\includegraphics[angle=270,width=\columnwidth]{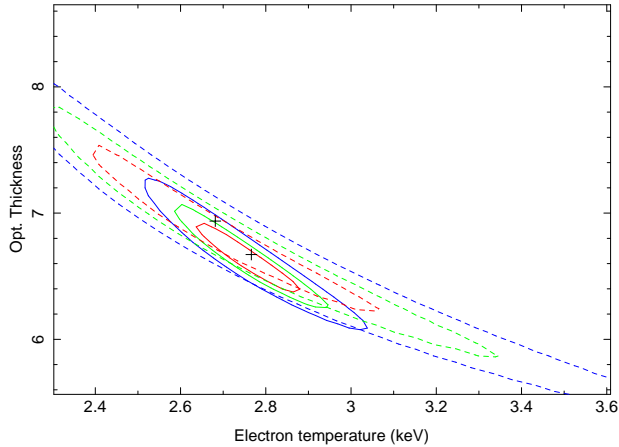}
\caption{Confidence contours for the $kT_e$ and $\tau$ parameters in the \diskbb+\comptt model fit for X-1, using only \xmm data (dashed) and  \xmm and \nus data (solid) of the second observation. The added value of NuSTAR data when it comes to constraining the electron temperature is evident.}
\label{fig:x1contours}
\end{figure}

\begin{deluxetable*}{lccccr}
\tablecolumns{6}
\tablewidth{0pc}
\tabletypesize{\scriptsize}
\tablecaption{{\bf X-1}: Best-fit Parameters for Some Common Spectral Models\label{tab:fitx1}}
\tablehead{			 & & \multicolumn{2}{c}{EPIC--pn only} & \multicolumn{2}{c}{pn, MOS2, FPM\{A,B\}} \\
\colhead{Parameter}&\colhead{Unit}&\colhead{Epoch 1}&\colhead{Epoch 2}&\colhead{Epoch 1}&\colhead{Epoch 2}}
\tablecomments{All uncertainties refer to single-parameter 90$\%$ confidence limits. \\
$^*$Values were fixed, or the parameter was unconstrained}
\startdata
\cutinhead{{\tt tbnew\_feo*(cutoffpl)} }
$n_{\rm H}$ & 10$^{22}\,{\rm cm^{-2}}$ & $0.267\pm0.006$ & $0.267\pm0.006$ & $0.265\pm0.006$ & $0.257\pm0.006$ \\
$N_{\rm cut}$ &  & $(7.9\pm0.1)\times10^{-4}$ & $(8.3\pm0.1)\times10^{-4}$ & $(7.9\pm0.1)\times10^{-4}$ & $(8.2\pm0.1)\times10^{-4}$ \\
$\Gamma$ &  & $2.00\pm0.02$ & $2.02\pm0.02$ & $1.97\pm0.03$ & $1.93\pm0.03$ \\
$E_{\rm cut}$ & keV & $500^{*}_{-192}$ & $500^{*}_{-181}$ & $73^{+62}_{-23}$ & $35^{+9}_{-6}$ \\
$\chi^2/{\rm dof}$ & & 1399/987 & 1379/968 & 2699/1952 & 2707/1939 \\
\cutinhead{{\tt tbnew\_feo*(diskbb+cutoffpl)} }
$n_{\rm H}$ & 10$^{22}\,{\rm cm^{-2}}$ & $0.27\pm0.02$ & $0.28\pm0.02$ & $0.27\pm0.01$ & $0.27\pm0.01$ \\
$N_{\rm dbb}$ &  & $11^{+4}_{-3}$ & $12^{+5}_{-3}$ & $10^{+3}_{-2}$ & $10\pm2$ \\
$T_{\rm in}$ & keV & $0.30\pm0.02$ & $0.29\pm0.02$ & $0.31\pm0.01$ & $0.31\pm0.01$ \\
$N_{\rm cut}$ &  & $(4.2\pm0.6)\times10^{-4}$ & $(4.7\pm0.6)\times10^{-4}$ & $(4.1\pm0.3)\times10^{-4}$ & $(4.2\pm0.3)\times10^{-4}$ \\
$\Gamma$ &  & $1.1^{+0.2}_{-0.3}$ & $1.2\pm0.2$ & $1.0\pm0.1$ & $1.0\pm0.1$ \\
$E_{\rm cut}$ & keV & $6^{+3}_{-2}$ & $9^{+5}_{-3}$ & $6.3^{+0.8}_{-0.7}$ & $5.8\pm0.6$ \\
$\chi^2/{\rm dof}$ & & 1037/985 & 1010/966 & 1998/1950 & 1986/1937 \\
\cutinhead{{\tt tbnew\_feo*(diskbb+comptt)} }
$n_{\rm H}$ & 10$^{22}\,{\rm cm^{-2}}$ & $0.27\pm0.02$ & $0.28\pm0.02$ & $0.28\pm0.01$ & $0.27\pm0.01$ \\
$N_{\rm dbb}$ &  & $23^{+9}_{-6}$ & $26^{+10}_{-7}$ & $25^{+7}_{-5}$ & $25^{+7}_{-5}$ \\
$T_{\rm in}$ & keV & $0.25\pm0.02$ & $0.25\pm0.02$ & $0.24\pm0.01$ & $0.24\pm0.01$ \\
$N_{\rm comp}$ &  & $(4.4\pm0.3)\times10^{-4}$ & $(4.2\pm0.4)\times10^{-4}$ & $(4.0\pm0.2)\times10^{-4}$ & $(4.3\pm0.2)\times10^{-4}$ \\
${kT_e}$ & keV & $2.4^{+0.3}_{-0.2}$ & $2.7^{+0.5}_{-0.3}$ & $2.8^{+0.2}_{-0.1}$ & $2.7\pm0.1$ \\
$\tau$ &  & $7.4\pm0.7$ & $6.9\pm0.7$ & $6.7\pm0.3$ & $6.8\pm0.3$ \\
$\chi^2/{\rm dof}$ & & 1033/985 & 1011/966 & 2020/1950 & 2013/1937 \\
\cutinhead{{\tt tbnew\_feo*(diskbb+comptt)} ($kT_{\rm e}=50$\,keV)}
$n_{\rm H}$ & 10$^{22}\,{\rm cm^{-2}}$ & $0.29\pm0.02$ & $0.28\pm0.02$ & $0.31\pm0.02$ & $0.31\pm0.02$ \\
$N_{\rm dbb}$ &  & $38^{+14}_{-9.6}$ & $36^{+13}_{-9}$ & $71^{+25}_{-18}$ & $98^{+40}_{-28}$ \\
$T_{\rm in}$ & keV & $0.22\pm0.01$ & $0.23\pm0.01$ & $0.19^{+0.01}_{-0.01}$ & $0.18\pm0.01$ \\
$N_{\rm comp}$ &  & $(2.6\pm0.2)\times10^{-5}$ & $(2.6\pm0.2)\times10^{-5}$ & $(3.2\pm0.2)\times10^{-5}$ & $(3.6\pm0.2)\times10^{-5}$ \\
$\tau$ &  & $0.78^{+0.05}_{-0.04}$ & $0.79\pm0.05$ & $0.63\pm0.02$ & $0.58\pm0.01$ \\
$\chi^2/{\rm dof}$ & & 1069/986 & 1028/967 & 2357/1951 & 2482/1938 \\
\cutinhead{{\tt tbnew\_feo*(powerlaw+kdblur2(1,reflionx))} }
$n_{\rm H}$ & 10$^{22}\,{\rm cm^{-2}}$ & $0.264^{+0.003}_{-0.01}$ & $0.265\pm0.009$ & $0.282\pm0.007$ & $0.282^{+0.004}_{-0.007}$ \\
$N_{\rm pow}$ &  & $4.16^{*}_{-0.05}\times10^{-4}$ & $(6.4\pm0.3)\times10^{-4}$ & $0^{*}$ & $0^{*}$ \\
$N_{\rm ref}$ &  & $1.5^{+0.3}_{-0.2}\times10^{-9}$ & $2^{+2}_{-1}\times10^{-8}$ & $(4.6\pm0.5)\times10^{-9}$ & $4.5^{+0.2}_{-0.5}\times10^{-9}$ \\
$A_{\rm Fe}$ &  & $20^{*}_{-1}$ & $5^{+8}_{-2}$ & $5.3^{+0.8}_{-0.5}$ & $5.9^{+0.6}_{-0.7}$ \\
$\Gamma$ &  & $1.82^{+0.03}_{-0.09}$ & $1.85^{+0.03}_{-0.02}$ & $1.65\pm0.04$ & $1.68^{+0.02}_{-0.04}$ \\
$X_{\rm i}$ &  & $(3\pm1)\times10^{3}$ & $(0.25^{+0.50}_{-0.01})\times10^{3}$ & $3.2^{+0.4}_{-0.3}\times10^{3}$ & $3.3^{+0.4}_{-0.2}\times10^{3}$ \\
$q$ &  & $6^{+3}_{-2}$ & $5^{+5}_{-3}$ & $7\pm3$ & $9.2^{+0.8}_{-0.3}$ \\
$R_{\rm in}$ &  & $1^{+7}_{*}$ & $1^{+6}_{*}$ & $1.32^{+0.29}_{-0.08}$ & $1.24\pm 0.03$ \\
$i$ & deg & $71\pm5$ & $81^{+5}_{-4}$ & $68^{+8}_{-16}$ & $75^{+1}_{-7}$ \\
$\chi^2/{\rm dof}$ & & 1011/973 & 973/953 & 2110/1877 & 2281/1934   
\enddata
\end{deluxetable*}

\fref{fig:x1overview} shows an overview of the spectral features of X-1. As can be seen in this plot, the spectrum did not change significantly between the two observations, either in the \xmm or in the \nus bands. We determined the cross-normalization constant between \nus and \xmm data to be $1.20\pm0.06$ for FPMA and $1.29\pm0.07$ for FPMB in the first observation, and $1.18\pm0.06$ FPMA and $1.25\pm0.07$ FPMB in the second. We measure an absorbed (0.3--10)\, keV luminosity of $(6.3\pm1.0)\times10^{39}\,{\rm erg\,s^{-1}}$ in the first observation ($\sim8.9\times10^{39}$ unabsorbed, assuming the best-fit {\tt diskbb+cutoffpl} model below) and $(6.6\pm1.0)\times10^{39}\,{\rm erg\,s^{-1}}$ in the second ($\sim9\times10^{39}$ unabsorbed). The corresponding absorbed 0.3--30\,keV luminosities are $(8.1\pm1.0)\times10^{39}$ and $(7.9\pm1.0)\times10^{39}\,{\rm erg\,s^{-1}}$, respectively. The spectral residuals with respect to the best-fit power law in the \xmm band are qualitatively  similar to the one reported from the 2006 October \xmm observation \citep{Dewangan+10}, associated with the low-flux state of this source, as opposed to the higher states where the soft excess is less prominent, as also shown in \fref{fig:x1overview}.

\subsubsection{Cutoff versus Reflection}
\begin{deluxetable}{lccr}
\tablecolumns{4}
\tablewidth{0pc}
\tabletypesize{\scriptsize}
\tablecaption{{\bf X-1}: Best-fit Parameters for \optxagnf, with the Data from All Instruments. \label{tab:fitoptx1}}
\tablehead{\colhead{Parameter}&\colhead{Unit}&\colhead{Epoch 1}&\colhead{Epoch 2}}
\tablecomments{All uncertainties refer to single-parameter 90$\%$ confidence limits. Note that for this model data were rebinned to 50 counts bin$^{-1}$ for \xmm, in order to reduce computation times during error bar calculations.}
\startdata
\cutinhead{{\tt tbnew\_feo*optxagnf} (norm fixed to 1)}
$n_{\rm H}$ & 10$^{22}$ & $0.27^{+0.01}_{-0.02}$ & $0.26\pm0.01$ \\
$M$ & solar & $93^{+17}_{-19}$ & $91^{+19}_{-14}$ \\
$\log L/L_{\rm Edd}$ &  & $-0.03^{+0.1}_{-0.07}$ & $-0.02^{+0.07}_{-0.08}$ \\
$R_{\rm cor}$ & rg & $66^{+6}_{-5}$ & $65\pm5$ \\
${kT_e}$ & keV & $2.0^{+0.2}_{-0.3}$ & $2.0\pm0.2$ \\
$\tau$ &  & $10^{+2}_{-1}$ & $10^{+1.5}_{-1.3}$ \\
$f_{\rm PL}$ &  & $0.6\pm0.1$ & $0.58^{+0.09}_{-0.12}$ \\
$\chi^2/{\rm dof}$ & & 1178/1190 & 1327/1220\\
\cutinhead{{\tt tbnew\_feo*optxagnf} (norm fixed to 2)}
$n_{\rm H}$ & 10$^{22}$ & $0.27^{+0.02}_{-0.01}$ & $0.26\pm0.01$ \\
$M$ & solar & $63^{+14}_{-10}$ & $65^{+13}_{-10}$ \\
$\log L/L_{\rm Edd}$ &  & $-0.16^{+0.08}_{-0.09}$ & $-0.18\pm0.08$ \\
$R_{\rm cor}$ & rg & $67^{+5}_{-6}$ & $65^{+5}_{-4}$ \\
${ kT_e}$ & keV & $2.0\pm0.2$ & $2.0\pm0.2$ \\
$\tau$ &  & $11^{+3}_{-2}$ & $10^{+1.9}_{-1.3}$ \\
$f_{\rm PL}$ &  & $0.65^{+0.09}_{-0.13}$ & $0.6\pm0.1$ \\
$\chi^2/{\rm dof}$ & & 1178/1190 & 1327/1220
\enddata
\end{deluxetable}

The first thing that becomes evident thanks to the \nus data is that the spectrum shows a clear cutoff above 10\,keV. As we mentioned earlier, hints of this cutoff are present in \xmm archival data of many ULXs, but this feature could be produced by a real cutoff or by relativistically smeared iron features. With the addition of \nus data this degeneracy is broken. We fitted the data with three models: (1) a power law with exponential cutoff (XSPEC model {\tt cutoffpl}), with and without an additional disk component modeled as a multicolor disk \citep[MCD; \diskbb;][]{Mitsuda+84} ; (2) \diskbb plus a Comptonization model \citep[\comptt;][]{comptt94} with the Comptonization seed photon temperature linked to the inner disk temperature for consistency; (3) a blurred reflection model obtained by convolving the {\tt reflionx} table \citep{reflionx05} with a Laor profile \citep{Laor91}, provided by the convolution model {\tt kdblur2} to account for general relativistic effects, following the method used by \citet{Walton+11} and \citet{CaballeroGarcia+10}. See \tref{tab:fitx1} for details.

While blurred reflection models and Comptonization/cutoff models yield similarly good fits in the \xmm band alone, they predict a completely different behavior around and above 10\,keV, as shown in \fref{fig:x1multi}. 
In reflection models, by adding \nus data we can find a decent nominal fit ($\chi^2/{\rm d.o.f.}\sim 1.08$ in the first observation, 1.18 in the second, see \tref{tab:fitx1}), but it is mostly due to the large number of \xmm spectral bins below 10\,keV. From the residuals in \fref{fig:x1multi} it is clear that the description of the spectrum is inadequate around and above 10\,keV. Even in the reflection-dominated regime where the power law normalization is zero and the downturn produced by the broadened iron line is maximum, the downturn is not sufficient to account for the very deep cutoff seen in \nus data, and the Compton ``hump'' produced by reflection clearly over predicts the spectrum above 10\,keV.

\subsubsection{Comparison of Comptonization Models}\label{sec:specx1comp}
\nus data enable us to obtain a much better constraint on the cutoff. This is shown in \fref{fig:x1contours}, where the contour levels between $kT_e$ and $\tau$ are shown with and without \nus data. It is clear from the contour plots that the addition of \nus data improves the constraint considerably. An alternative way to show the poor constraint given by \xmm data alone is to fix the electron temperature of \comptt ($kT_{\rm e}$) to 50\,keV, and fit the data. If we take \xmm data only, the fit deteriorates ($\Delta\chi^2\sim20$ in the first observation, $\sim30$ in the second for  {$\Delta {\rm d.o.f.=1}$) but we can still recover an overall acceptable fit ($\chi^2/{\rm d.o.f.}\sim 1.08$ in the first, 1.06 in the second) and even obtain a compatible value of the disk temperature (\tref{tab:fitx1}). \xmm response drops and data have only a few points around 10\,keV, where the constraint on the cutoff is set, and small systematic errors in the instrument response can influence the fit. The difference between the two models disappears if one discards the last 20 bins of the spectrum. With the addition of \nus data this is not true anymore, and in fact the fit deteriorates further ($\chi^2/{\rm d.o.f.}> 1.2$) and the disk temperature assumes incompatible values (see \tref{tab:fitx1}). 

From \tref{tab:fitx1} and \fref{fig:x1multi} it is also clear that the \comptt model gives a slightly worse fit than the simple {\tt cutoffpl} model, with a high-energy slope visibly steeper than what \nus data show. This fact indicates that a single-temperature Comptonization model is probably not sufficient to describe the data. We make use of the \optxagnf model \citep{optxagn12}, which is a phenomenological model that represents the evolution of the \dkbbfth model \citep{dkbbfth06} often used for ULXs in the past \citep[e.g.][]{Gladstone+09,Walton+11}. \optxagnf, originally developed for active galactic nuclei (AGN), tries to balance in a self-consistent way the optically thick emission from the disk, a low-temperature Comptonization component originating from the inner part of the disk, and a second, hot Comptonization component with cutoff above 100\,keV produced by a hot corona. With respect to the \dkbbfth model, \optxagnf adds a second Comptonizing component while maintaining the possibility of hiding the underlying disk emission below a corona that covers the disk and is powered by it. The latter was the reason \dkbbfth was used in the past. Moreover, \optxagnf has superior computational stability and a more convenient choice of parameters, using the expected mass, spin and luminosity of the BH instead of a generic normalization parameter linked to the position of the inner disk (in fact, in this model the normalization factor should normally be frozen to 1, but see below). 

This model, however, has nine  free parameters (mass, spin, luminosity, photon index and normalization of the hot Comptonizing component, optical thickness and temperature of the cold Comptonizing component, radius of this cold component, and outer radius of the disk), and therefore it is able to yield many different solutions for a given spectrum.  
We therefore restrict the parameter space by fixing some  of them to reasonable values and discuss the results obtained with this approach, with the obvious associated caveats. 
A discussion of the full range of scenarios that this model can describe is beyond the scope of this work and will be discussed in a future paper.

Our \nus data show an excess with respect to a single-temperature Comptonization model, but do not show the  plateau at high energies that has been observed for example in the bright AGN \citep{optxagn12}. We therefore fix the power law index of the hot electrons, $\Gamma$, to 2, a typical value observed in BH power law spectra, and we free only its normalization factor, $f_{\rm PL}$. We also fix the outer disk to $10^5\,\rg$, the spin parameter $a$ to 0 and, as is prescribed in the documentation, the normalization factor to 1. Because this model does not take into account the inclination and assumes an observing angle of 60\degree, and the norm is proportional to $\cos i / \cos 60\degree$, we also fitted the data with the norm fixed to 2 (source seen face-on) to evaluate whether a change in this parameter could dramatically affect the results. 

We summarize the best fit in this reduced parameter space in \tref{tab:fitoptx1}. 
In both observations this model yields an intriguing result: under the above assumptions ($a$=0, $\Gamma=2$), and with both normalizations, the spectrum seems to be well described by a quite massive ($\sim$70--90\msun) BH, accreting close to (or slightly above) Eddington, with a large corona reaching $\sim60\,\rg$. The fraction of energy that is reprocessed from the hot part of the corona is about 60\%, while the rest is reprocessed by the cold and optically thick part. As expected, fixing the norm to 2 has the effect of lowering both the mass of the BH and the luminosity, but the rest of parameters do not change significantly.

All of the above models leave some residuals around 1\,keV and below. They appear very similar in all fits, indicating that they are independent from the particular continuum model used. Similar residuals are often observed in ULXs \citep[see, e.g.,][]{Soria+04,GoncalvesSoria06,Gladstone+09, CaballeroGarcia+10}. We tested the improvement of the fit with the addition of a MEKAL component \citep{MeweGronenshild81} to the {\tt cutoffpl} and {\tt diskbb+cutoffpl} models. We failed to obtain a good fit in the first case, while in the second we found a general improvement of the fit($\Delta\chi^2\sim50$), with a MEKAL temperature of about 1\,keV and the abundances fixed to the standard values. As the diffuse emission from the NGC 1313 galaxy is negligible, this might indicate the presence of emission from a hot medium close to the source. 

\subsection{NGC~1313 X-2}\label{sec:specx2}
\begin{figure}
\centering
\includegraphics[angle=270,width=\columnwidth]{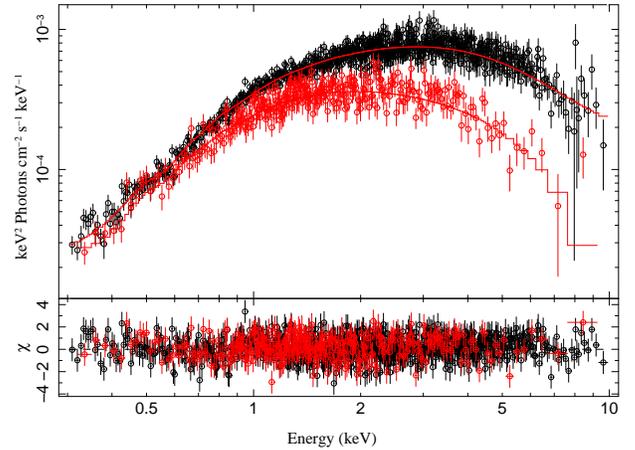} 
\caption{\xmm unfolded spectra of NGC~1313 X-2 during the two observations. Black points are EPIC-pn and red points are EPIC-MOS1 data, due to the source being in the gap of pn in the second observation. The best-fit slim-disk model for the two observations with {\tt modelID} set to 4 is superimposed.}.
\label{fig:x2overview}
\end{figure}

\begin{deluxetable}{lccr}
\tablecolumns{4}
\tablewidth{0pc}
\tabletypesize{\scriptsize}
\tablecaption{{\bf X-2}: Best-fit Parameters for Some Models, with the Data from \xmm Only. \label{tab:fitoptx2}}
\tablehead{\colhead{Parameter}&\colhead{Unit}&\colhead{Epoch 1}&\colhead{Epoch 2}}
\tablecomments{All uncertainties refer to single-parameter 90$\%$ confidence limits.}
\startdata
\cutinhead{{\tt tbnew\_feo*(cutoffpl)} }
$n_{\rm H}$ & 10$^{22}\,{\rm cm^{-2}}$ & $0.23\pm0.02$ & $0.29\pm0.02$ \\
$N_{\rm cut}$ &  & $(7.6\pm0.2)\times10^{-4}$ & $(5.8\pm0.2)\times10^{-4}$ \\
$\Gamma$ &  & $0.9\pm0.1$ & $1.5\pm0.2$ \\
$E_{\rm cut}$ & keV & $2.5\pm0.2$ & $2.5^{+0.4}_{-0.3}$ \\
$\chi^2/{\rm dof}$ & & 776/833 & 613/528\\
\cutinhead{{\tt tbnew\_feo*(diskbb)} }
$n_{\rm H}$ & 10$^{22}\,{\rm cm^{-2}}$ & $0.153^{+0.006}_{-0.005}$ & $0.142\pm0.008$ \\
$N_{\rm dbb}$ &  & $0.056\pm0.003$ & $0.106\pm0.009$ \\
$T_{\rm in}$ & keV & $1.21\pm0.02$ & $0.86\pm0.02$ \\
$\chi^2/{\rm dof}$ & & 959/834 & 833/529\\
\cutinhead{{\tt tbnew\_feo*(diskpbb)} }
$n_{\rm H}$ & 10$^{22}\,{\rm cm^{-2}}$ & $0.27\pm0.02$ & $0.320^{+0.008}_{-0.011}$ \\
$N_{\rm dbb}$ &  & $0.010\pm0.002$ & $0.0059^{+0.0013}_{-0.0008}$ \\
$T_{\rm in}$ & keV & $1.56\pm0.06$ & $1.27\pm0.05$ \\
$p$ &  & $0.58\pm0.01$ & $0.500^{+0.006}_{*}$ \\
$\chi^2/{\rm dof}$ & & 776/833 & 611/528\\
\cutinhead{{\tt tbnew\_feo*(diskbb+cutoffpl)} }
$n_{\rm H}$ & 10$^{22}\,{\rm cm^{-2}}$ & $0.22^{+0.03}_{-0.02}$ & $0.24^{+0.04}_{-0.02}$ \\
$N_{\rm dbb}$ &  & $0.7$  (unconstr.) & $1.7^{+1.6}_{-0.8}$ \\
$T_{\rm in}$ & keV & $0.4^{+0.1}_{-0.2}$ & $0.38^{+0.09}_{-0.10}$ \\
$N_{\rm cut}$ &  & $6^{+2}_{-3}\times10^{-4}$ & $3^{+3}_{-2}\times10^{-4}$ \\
$\Gamma$ &  & $0.5^{+1.3}_{-1}$ & $-0.3^{+1.4}_{-1.2}$ \\
$E_{\rm cut}$ & keV & $2.0\pm0.5$ & $1.2^{+2.2}_{-0.2}$ \\
$\chi^2/{\rm dof}$ & & 773/831 & 607/526\\
\cutinhead{{\tt tbnew\_feo*(diskbb+comptt)} }
$n_{\rm H}$ & 10$^{22}$ & $0.183\pm0.008$ & $0.50^{+0.05}_{-0.04}$ \\
$N_{\rm dbb}$ &  & $0.20^{+0.08}_{-0.06}$ & $4^{+12}_{-3}\times10^{4}$ \\
$T_{\rm in}$ & keV & $0.81 \pm 0.1$ & $0.07\pm0.01$ \\
$N_{\rm comp}$ &  & $(1.2\pm0.4)\times10^{-4}$ & $(2.6^{+0.7}_{-0.4})\times10^{-4}$ \\
${kT_e}$ & keV & $2.0^{+0.1}_{*}$ & $2.0^{+0.1}_{*}$ \\
$\tau$ &  & $7.6^{+1.7}_{-0.8}$ & $5.0^{+0.1}_{-0.2}$ \\
$\chi^2/{\rm dof}$ & & 811/831 & 661/526\\
\cutinhead{{\tt tbnew\_feo*optxagnf} }
$n_{\rm H}$ & 10$^{22}\,{\rm cm^{-2}}$ & $0.18\pm0.01$ & $0.17^{+0.04}_{-0.02}$ \\
$M$ & \msun & $21^{+4}_{-3}$ & $32^{+27}_{-9}$ \\
$\log L/L_{\rm Edd}$ &  & $0.29^{+0.06}_{-0.07}$ & $-0.1^{+0.1}_{-0.2}$ \\
$R_{\rm cor}$ & rg & $39^{+61}_{-16}$ & $92^{+8}_{-68}$ \\
${kT_e}$ & keV & $1.12^{+0.13}_{-0.08}$ & $0.87^{+0.06}_{-0.07}$ \\
$\tau$ &  & $16^{+4}_{-3}$ & $13^{+7}_{-1}$ \\
$\chi^2/{\rm dof}$ & & 559/599 & 443/422
\enddata
\end{deluxetable}

\begin{deluxetable*}{lccccccc}
\tablecolumns{8}
\tablewidth{0pc}
\tabletypesize{\scriptsize}
\tablecaption{{\bf X-2}: Best-fit parameters for X-2, with the \slimdisk model, and the mass tied between the two observations. Subscripts 1 and 2 refer to the two observations.\label{tab:x2_kaw}}
\tablehead{\colhead{\tt modelID} & \colhead{$\alpha$ } &\colhead{$n_{\rm H,1}$ (${\rm cm^{-2}}$) }    & \colhead{$n_{\rm H,2}$ (${\rm cm^{-2}}$) }    &
 \colhead{$M$ (\msun)}    & \colhead{$\dot{M}_1$ ($L_{\rm Edd}/c^2$) } & \colhead{$\dot{M}_2$ ($L_{\rm Edd}/c^2$) }  & \colhead{$\chi^2/\mathrm{dof}$}}
\tablecomments{The mass was tied in the two observations. All uncertainties refer to single-parameter 90$\%$ confidence limits. {\tt modelID}=4 means that we are modeling a slim disk plus Comptonization. With {\tt modelID}=7, we are also adding relativistic corrections. Being the grid of the \slimdisk model quite sparse, the errors on the parameters are typically inside the range between a value and the following in the grid. For this reason, one should use these uncertainties with some caution. }
\startdata
 7 & $0.0100^{+0.0006}_{*}$&$0.296\pm0.006$			& $0.23\pm0.01$& $21.4\pm0.8$		& $30\pm2$	  & $12.2^{+0.4}_{-0.3}$		& 1521/1422 \\ 
 4 & $0.11^{+0.04}_{-0.02}$& $0.265^{+0.007}_{-0.006}$ & $0.23\pm0.01$ & $36^{+2}_{-1}$ & $11.5^{+0.4}_{-0.5}$ & $5.1\pm0.2$ & 1456/1422
\enddata
\end{deluxetable*}

As described in \sref{sec:nusproc}, we did not use \nus data for the analysis of X-2. 
\fref{fig:x2overview} shows the shape of the \xmm spectrum of X-2 in the two epochs. The flux and overall shape of the spectrum changed considerably between the two observations, as also did the timing behavior (see \sref{sec:x2timing}). We measure a 0.3--10\,keV absorbed luminosity of $(4.6\pm0.6)\times10^{39}\,{\rm erg\,s^{-1}}$ in the first epoch and $(2.2\pm0.6)\times10^{39}\,{\rm erg\,s^{-1}}$ in the second epoch.

Spectral fits with several models are presented in \tref{tab:fitoptx2}. The spectrum is reasonably well described by an absorbed cutoff power law in both epochs, but the values of the spectral index (down to 0.9 in one case) are very different from what would be expected by Comptonization, the main process known to produce this kind of spectral shape, that instead yields spectral indices between 1.5 and 3. The addition of a disk component to the cutoff power law barely improves the fit ($\Delta\chi^2\sim 3$ in the first observation, and $\sim6$ in the second, for 2 fewer degrees of freedom, dof). Similarly, a {\tt diskbb+comptt} model does not improve the fit with respect to the {\tt cutoffpl} model. 

The spectral shape is clearly not well described by a standard MCD (XSPEC model \diskbb), but it is well modeled by a so-called $p$-free disk \citep[\diskpbb,][]{Mineshige+94, Kubota+05}. When the accretion rate is high, it is expected that the structure of the disk deviates considerably from the standard \citet{SS73} thin disk. In this model, the radial dependency of the disk temperature is parameterized with $T\propto r^{-p}$, where $p$ is different from the $3/4$ value used in the thin disk. The $p$-disk would recover the standard thin disk if $p=0.75$. For $p<0.75$ the temperature profile is affected by advection. At $p$=0.5, advection dominates and the disk is a so-called {\em slim disk} \citep{Abramowicz+88,WataraiFukue99}.  

The $p$-free model seems to yield a very good fit for both epochs, with values of $p$ very close to the slim disk regime. The amount of advection is closer to the slim disk regime in the fainter observation. This behavior has been reported for this and other ULXs in the past \citep{Mizuno+07, Middleton+11M33, Straub+13}. The deviation of $p$ from 0.5 in the brightest observation might imply some reprocessing of the disk emission, for example by a corona. 

The fact that these two observations have such different fluxes and spectral shapes gives us the opportunity to jointly fit the data with more physically-motivated models that would be difficult to constrain with  single observations, and try to obtain an estimate on the mass of the source.

With this goal in mind, we used an advanced slim disk model, implemented by \citet[hereafter \slimdisk]{Kawaguchi03}. In this local XSPEC table model (used in the past for fitting ULX spectra, see, e.g. \citealt{Vierdayanti+06,Godet+12}), mass $M$ (in \msun) and accretion rate $\mdot$ are the only physical parameters. \mdot is calculated in units of $L_{\rm Edd}/c^2$, where $L_{\rm Edd}$ is the Eddington luminosity. Since $L\simeq \eta \mdot c^2$, where $\eta$ is the efficiency, then the value of \mdot corresponding to the Eddington luminosity is $1/\eta\sim 16$, assuming the efficiency for a Schwarzschild BH calculated by using a pseudo-Newtonian potential \citep[see, e.g.,][]{Ebisawa+03}. Comptonization from a corona, gravitational redshift and transverse Doppler effect are included self-consistently, but there is no observing angle dependence, as the source is assumed to be face-on. Ideally, this model provide a unique value of  the mass given the mass accretion rate, or vice versa. The choice of the spectral model (slim disk alone, with an additional thermal component or Comptonization, with or without relativistic effects) to include in the computation is done by switching the values of the {\tt modelID} parameter. For our purposes, we are interested in the treatment of a slim disk with Comptonization, and with or without relativistic effects (i.e. using {\tt modelID} equal to 4 or 7). The disk viscosity parameter $\alpha$, the only non-observable quantity, can also be set. 

We tied almost all parameters of the model in the two epochs, leaving only the mass accretion rate $\dot{M}$ free to vary between them; we fixed the normalization to $(10\mathrm{kpc}/d)^2=5.86\times10^{-6}$, assuming $d=4.13$\,Mpc. 

We initially fixed the {\tt modelID} to 7, meaning that we used all corrections for gravity and Comptonization. We tried different values for the $\alpha$ parameter, and found that the spectral shape was best described by $\alpha\sim0.01$, i.e., the lower limit of this table for the viscosity parameter. Even with these very restrictive assumptions, the model was able to fit the data quite well (see Table~\ref{tab:x2_kaw}). 

By taking out the relativistic corrections, namely changing the {\tt modelID} to 4, we were again able to fit the data fairly well. In this case, there was need for a higher viscosity in order to reproduce the curvature of the spectrum. As a result the measured values of the mass are higher, but always in the range of StMBHs.

In addition to the \slimdisk model, we used the aforementioned \optxagnf. We started by fitting the model to the single observations, similarly to what was done for X-1 (best-fit results in \tref{tab:fitoptx2}), this time fixing the power law index to 2.2 (by analogy with high-accretion rate Seyfert galaxies) and the fraction of emission in the hot corona to 0.3. Then we fitted together the two observations. Given the complexity of this model, we used it to obtain estimates on the most likely source parameters by fixing the mass, the photon index of the hot electrons $\Gamma$ and the spin $a$ to discrete values (10, 30, 60, 90\,\msun for $M$, 1.8, 2 and 2.2 for $\Gamma$, 0 and 0.998 for $a$) and freeing the fraction of hot power law emission, the optical thickness and the cold corona temperature. The norm was fixed to 1 and then to 2, because of the arguments considered in \sref{sec:specx1comp}. For {\tt norm}=1, we found the best fit ($\chi^2/{\rm dof} = 1.001$) for $M=30\msun$ and $\Gamma=2.2$. The fit with $M=10\,\msun$ was always unacceptable ($\chi^2/{\rm dof} \gtrsim 1.7$). For $M=60\msun$ we could obtain an acceptable fit ($\chi^2/{\rm dof} \lesssim 1.05$) only for $\Gamma=2$. In all other cases either the fit was worse, or one or more parameters reached their hard limits, implying a non-ideal regime of the model. Varying the spin from 0 to 0.998 did not change the results dramatically, with $\tau$, $kT_{\rm e}$ and $f_{\rm PL}$ compensating for most of the change in spectral shape. As before, fixing the norm to 2 lowered the estimate on the mass, permitting to obtain decent fit values ($\chi^2/{\rm dof} = 1.06$) also for $M=10\msun$ and $\Gamma=$1.8--2.

As a bottom line, the favored interpretation, from both \optxagnf and \slimdisk, seems to be a StMBH (up to $\sim$50\,\msun) accreting around Eddington, or transitioning between a super-Eddington and a sub-Eddington regime. The emission from the cold and thick corona given by \optxagnf, extending over a large region of the inner disk, does not differ substantially from the bloated disk described in the \slimdisk model, and it is thus not surprising that the two models produce similar results. 

The same caveats discussed for X-1 apply here: this result is model-dependent and based on the assumptions we made about the parameters; only further investigation using more observations with different spectral states will tell if the constraints on the mass are robust. 

%
%

\section{Timing analysis}\label{sec:timing}
We extracted filtered event lists for both ULXs from all datasets and produced lightcurves cleaned from gaps and periods of increased background activity. These data were then processed with the following timing analysis techniques.

\subsection{rms variability}
The first variability test we used on our data is the normalized {\em excess variance} test \citep{Edelson+90,Vaughan+03}. Let $S$ be the intrinsic variance of the source signal (as calculated from the lightcurve), $\sigma_{i}$ the standard error on the $i$th bin of the lightcurve (calculated from Poissonian statistics) and $\bar{\sigma}$ the mean standard error, $\bar{I}$ the mean counts per bin in the lightcurve. The excess variance is then simply $S - \bar{\sigma}^2$; we normalize it as follows:
\begin{equation}
F_{\rm var} = \sqrt{\frac{S - \bar{\sigma^2}}{\bar{I}^2}}.
\end{equation}
$F_{\rm var}$ has the advantage of being a linear quantity, and thus yields a measure of the intrinsic root mean square (rms) variability of the source. 
The error we quote is the one derived in \citet{Vaughan+03}.

\subsection{Power Density Spectrum}
For each lightcurve, we extracted a power density spectrum (PDS), the normalized square modulus of the Fourier Transform (see \citealt{VDK89} for an extensive review of the methods used in the following). We used the \citet{Leahy+83} normalization, so that the PDS has a white noise level of 2. Dead time effects can safely be ignored due to the very low count rates of the sources analyzed. 

This timing analysis is very sensitive to data gaps in lightcurves, which produce low-frequency noise and spikes in the PDS. \nus data, because of the very short orbital period of the satellite ($\sim 90$\,minutes) and the position of the source, have about $\sim 30$\,minutes of occultation every orbit. Moreover, both \xmm and \nus data have other gaps due to, for example, the filtering of periods of high background activity. As a strategy in our analysis, we decided to fill gaps of very short length (several seconds) with white noise at the average count rate in the nearby 4000s of data. We verified that, due to the very low count rate, this did not produce any spurious features in the spectrum. Data chunks with longer gaps, such as occultation periods, were simply ignored. This also limits the maximum length of single fast Fourier Transoforms for \nus data to less than $\sim 1$hr, while there is no such constraint for \xmm. 

We used different rebinning factors in order to look for features with different spectral width. Following \citet{Barret+12}, we used maximum-likelihood fitting to evaluate features in cases where the rebinning was not sufficient to attain the Gaussian regime. The maximum frequency investigated was $512$ Hz, to include possible high-frequency quasi-periodic oscillations (QPOs) as often observed in BH sources \citep[see][for reviews]{Remillard+06,Belloni+12}. The minimum frequency was the inverse of the length of each analyzed chunk with no gaps. For \nus this was limited to $\sim0.3$\,mHz, while for \xmm data $\sim0.1$\,mHz. 

\subsection{NGC~1313 X-1}\label{sec:x1timing}
The PDS of NGC~1313 X-1 is almost featureless. The Kolmogorov-Smirnov test, calculated from the lightcurve at different bin times, does not detect any variability and we find no significant detections of QPOs or low-frequency noise in the PDS. $F_{\rm var}$ is consistent with 0. This source  historically showed variability only in its brighter states. \citet{Dewangan+10} studied the relation between variability and spectral states in X-1 and our timing results are, together with our spectral results (\sref{sec:specx1}) compatible with what they call the ``low-flux'' state. 

\subsection{NGC~1313 X-2}\label{sec:x2timing}
The behavior of this source is quite interesting from the timing point of view. The results of the timing analysis are shown in \fref{fig:x2timing}. 
The change in spectral shape observed in this source (\sref{sec:specx2}) is also reflected in the timing properties. As \fref{fig:x2timing} shows, in the observation with higher flux the power spectrum shows low-frequency variability. The overall rms \`a la \citet{Vaughan+03} is $F_{\rm var} = 13.6(7)\%$ in the first observation, and very low, consistent with 0, in the second. The PDS of the first observation shows a red-noise component but no significant QPOs. This variability increases with energy (see\fref{fig:x2timing}). This is probably a hint of what models are more likely to describe the spectra. In fact, if a pure (slim) disk was responsible for this emission, the higher-energy variability would correspond to the part of the disk {\em closest} to the BH, where variability timescales should be {\em faster}, surely well above 1\,Hz. But the PDS shows that this variability is mostly at low frequencies ($<1$\,mHz). This gives support to a geometry where instead the source of variability is the Comptonizing medium, whose contribution increases with energy and whose timescales are not necessarily linked to the timescales in the inner disk.

\begin{figure*}
\centering
\includegraphics[width=6.6in]{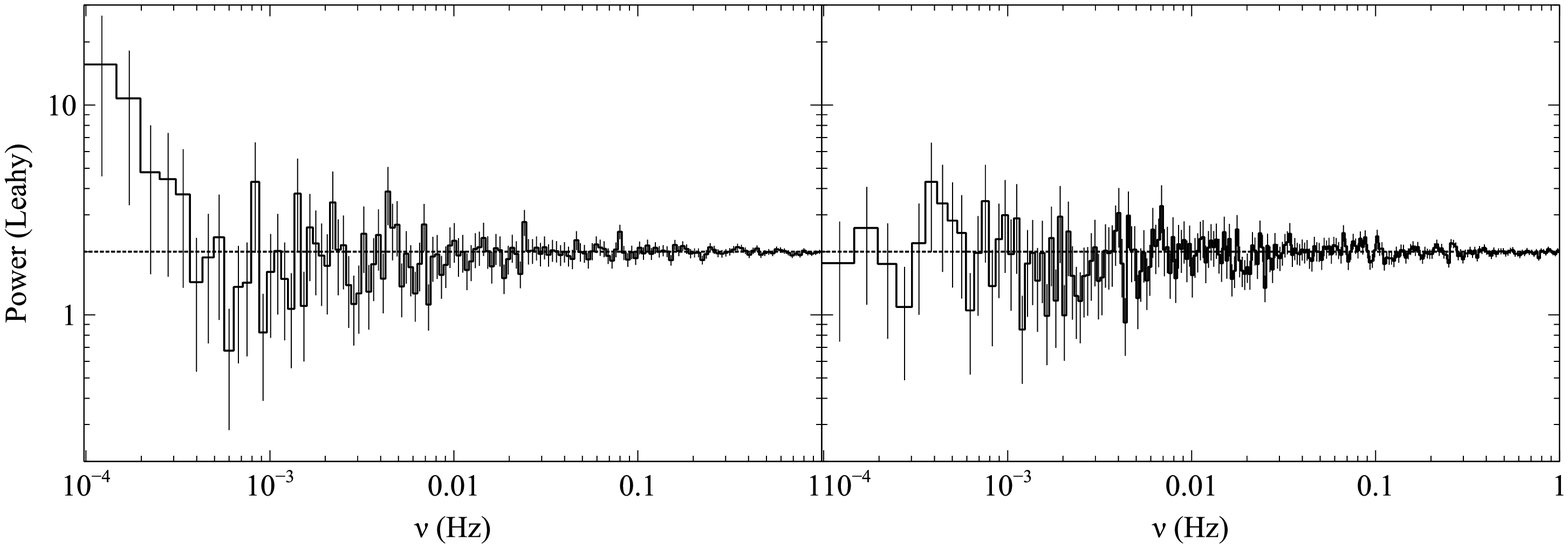} \\
\includegraphics[width=6.6in]{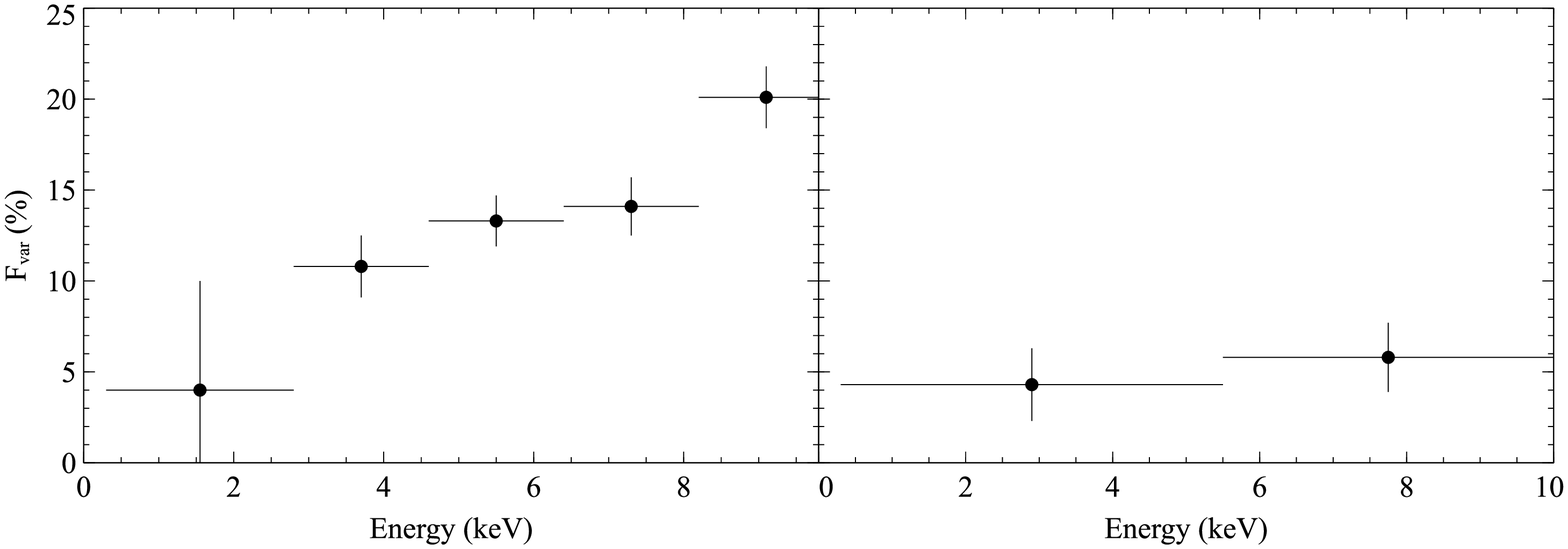}
\caption{NGC~1313 X-2, from top to bottom: PDS with geometric rebinning (bin factor 1.03), and rms  spectrum calculated from the normalized excess variance $F_{\rm var}$ (see \sref{sec:x2timing} for  details). From left to right, first and second observations. We did not include the full range of frequencies investigated for clarity. Dotted lines in the PDS show the white noise level of 2. Errors in the PDS are calculated as $2 / \sqrt{MW}$, with $M$ the number of averaged PDSs and $W$ the number of averaged nearby bins. We only use \xmm data due to the poor quality of \nus data for this source (see \sref{sec:nusproc}). In the rms spectrum we only plot points where $F_{\rm var} > 0$. According to the PDS and the rms spectrum significant variability is only present in the first observation.}
\label{fig:x2timing}

\end{figure*}

%
%

\section{Discussion}\label{sec:discussion}
In this paper we present the first \nus+\xmm results on the two ULXs in NGC~1313. \nus data have proven particularly useful for X-1, where the data above 10\,keV clearly show a cutoff that was not well constrained by \xmm (see \fref{fig:x1contours}). In X-2, due to the soft spectrum and the unfavorable position in the field of view, \nus data are not as decisive, but \xmm data are sufficient to perform high-quality spectral and timing analysis below 10\,keV.

\subsection{X-1}
The results obtained for this source thanks to \nus data represent a new landmark in the understanding of ULX physics. Before \nus was launched, ULX spectra had been studied in detail only below 10\,keV. At least two different models were previously able to describe the spectral energy distribution: a reflection-dominated regime where the downturn is produced by a very strong and broadened iron line, and several combinations of MCDs (or other kinds of soft excess models) and low-temperature Comptonized emission cutting off slightly below 10\,keV. 

We show also in this paper that, even with a $\sim100$\,ks pointing, the \xmm spectrum alone is not sufficient to constrain the cutoff
. Also using a reflection model gives a very nice description of the spectrum with low residuals and good $\chi^2$ in the \xmm band. With \xmm alone, the spectrum might describe a standard low-hard state of a quite massive BH with a strong Comptonized component from a hot and optically thin medium, a reflection dominated state where the underlying power law is not  observable, or a soft disk component and a reprocessed component that cuts off around 10\,keV.

The addition of \nus data removes this degeneracy. 
In the \nus band, the spectrum shows a very clear cutoff around 10\,keV, similar in character to that expected from Comptonization by a cold, thick medium, but slightly less steep. The quality of \nus data is such that we can put tight constraints on the cutoff energy as shown in \fref{fig:x1contours} and \tref{tab:fitx1}, and, as a result, the significance of a low-temperature disk component detected by \xmm. The presence of a low-temperature, optically thick Comptonized component suggests that we are observing accretion at high Eddington fractions that make the geometry of the system deviate substantially from the standard picture valid for lower luminosity BHs and confirms previous, albeit much less constraining, observations \citep[e.g.][]{Stobbart+06}.

By going into more detail and fitting the \optxagnf phenomenological model that includes a color-corrected MCD plus a two-component corona composed of a cold, optically thick medium and a second, hot and optically thin one, we obtain an interesting result: X-1 would be a quite massive StMBH of about 70--100\,\msun, accreting close to Eddington, with a large, cold corona covering a significant part of the inner disk. This is in agreement with the lack of signatures of strong outflows that should be associated with highly super-Eddington accretion \citep[see, e.g.,][]{Poutanen+07}, such as photo-ionized bubbles (that are seen instead for other sources, e.g., in \citealt{PakullMirioni02,Ramsey+06}) or  discrete atomic features in their high-energy spectra that could be associated with either iron emission or absorption from a wind \citep{Walton+12}. An alternative explanation is that these winds are not pointing towards the observer, and the source is observed almost face-on (this would also agree with the lack of variability; see, e.g., \citealt{Middleton+11, Sutton+13}).

To summarize, this source is clearly not accreting in a standard BH hard or soft state, as is shown by the absence of a power law and a spectrum not dominated by disk emission, and hence the high luminosity and the cold inner disk are not indicative of the mass. The spectral shape is instead well described by what is generally associated with accretion close to Eddington, that is an optically thick corona covering the inner part of the disk, and the energetics of the system points towards the high-end of the StMBH population.

\subsection{X-2}

We caught a large spectral variation in X-2 that is extremely interesting both for the rapidity (one week) of the change and for its characteristics. The higher state is the one with the higher variability. This spectral behavior is reminiscent of the hard state of known BHs, where there is a linear correlation between rms variability and flux (see, e.g., \citealt{UttleyMcHardy01}; \citealt{McHardy10} for a review).
Nonetheless, the shape of the spectra  do not match the general picture of spectra in the hard state, where a prominent power law component is usually present \citep{Done+07}. In our spectra the high-energy component drops off very quickly below or around 10 keV.

The spectrum of the source is instead well described by a StMBH with an advection-dominated disk, or {\em slim disk}, accreting around the Eddington limit. By linking the mass between the two observations and fitting the \slimdisk model, and independently by using the \optxagnf model, we are able to obtain an estimate of the mass of the BH around 25\,\msun and a luminosity that is shifting from super-to sub-Eddington.

Even if Comptonization were not required by the spectrum, the fact that the most variability comes from higher energies, as shown in \fref{fig:x2timing}, gives support to models including this component. A simple {\em slim} disk would be able to yield this variation of rms with energy but we would expect the highest energies to be produced in the region of the disk closer to the BH, where variability is faster. This is in contradiction to the very low frequencies we observe in the PDS, and makes us favor an interpretation where variability comes instead from the corona, whose relative contribution to the spectrum is more important at high energies. 

\section {Acknowledgements}
M.B. wishes to acknowledge the support from the Centre National d'\'Etudes Spatiales (CNES). 
This work was supported under NASA Contract No. NNG08FD60C, and
made use of data from the {\it NuSTAR} mission, a project led by
the California Institute of Technology, managed by the Jet Propulsion
Laboratory, and funded by the National Aeronautics and Space
Administration. We thank the {\it NuSTAR} Operations, Software and
Calibration teams for support with the execution and analysis of
these observations.  This research has made use of the {\it NuSTAR}
Data Analysis Software (NuSTARDAS) jointly developed by the ASI
Science Data Center (ASDC, Italy) and the California Institute of
Technology (USA). This work also makes use of observations obtained with \xmm, an ESA science mission with instruments and contributions directly funded by ESA Member States and NASA, and of observations made by the {\it Chandra X-ray Observatory}. For timing analysis and plotting, a set of Python codes making use of the NumPy and Scipy libraries was used. For some plots, we used the Veusz software.
The authors wish to thank Olivier Godet and Chris Done for interesting discussions, and the referee Matt Middleton, whose comments and suggestions substantively improved the quality of the manuscript.

\bibliographystyle{apj}
\bibliography{papers2,additional}

\begin{thebibliography}{}
\expandafter\ifx\csname natexlab\endcsname\relax\def\natexlab#1{#1}\fi

\bibitem[{Abramowicz {et~al.}(1988)Abramowicz, Czerny, Lasota, \&
  Szuszkiewicz}]{Abramowicz+88}
Abramowicz, M.~A., Czerny, B., Lasota, J.~P., \& Szuszkiewicz, E. 1988, ApJ,
  332, 646

\bibitem[{Arnaud(1996)}]{Arnaud96}
Arnaud, K.~A. 1996, ADASS, 101, 17

\bibitem[{Barret \& Vaughan(2012)}]{Barret+12}
Barret, D., \& Vaughan, S. 2012, ApJ, 746, 131

\bibitem[{Belloni {et~al.}(2012)Belloni, Sanna, \& Mendez}]{Belloni+12}
Belloni, T.~M., Sanna, A., \& Mendez, M. 2012, MNRAS, 426, 1701

\bibitem[{Caballero-Garcia \& Fabian(2010)}]{CaballeroGarcia+10}
Caballero-Garcia, M.~D., \& Fabian, A.~C. 2010, MNRAS, 402, 2559

\bibitem[{Dewangan {et~al.}(2010)Dewangan, Misra, Rao, \&
  Griffiths}]{Dewangan+10}
Dewangan, G.~C., Misra, R., Rao, A.~R., \& Griffiths, R.~E. 2010, MNRAS, 407,
  291

\bibitem[{Done {et~al.}(2012)Done, Davis, Jin, Blaes, \& Ward}]{optxagn12}
Done, C., Davis, S.~W., Jin, C., Blaes, O., \& Ward, M. 2012, MNRAS, 420, 1848

\bibitem[{Done {et~al.}(2007)Done, Gierli{\'n}ski, \& Kubota}]{Done+07}
Done, C., Gierli{\'n}ski, M., \& Kubota, A. 2007, Astron Astrophys Rev, 15, 1

\bibitem[{Done \& Kubota(2006)}]{dkbbfth06}
Done, C., \& Kubota, A. 2006, MNRAS, 371, 1216

\bibitem[{Ebisawa {et~al.}(2003)Ebisawa, {\.Z}ycki, Kubota, Mizuno, \&
  Watarai}]{Ebisawa+03}
Ebisawa, K., {\.Z}ycki, P., Kubota, A., Mizuno, T., \& Watarai, K.-y. 2003,
  ApJ, 597, 780

\bibitem[{Edelson {et~al.}(1990)Edelson, Krolik, \& Pike}]{Edelson+90}
Edelson, R.~A., Krolik, J.~H., \& Pike, G.~F. 1990, ApJ, 359, 86

\bibitem[{Farrell {et~al.}(2009)Farrell, Webb, Barret, Godet, \&
  Rodrigues}]{Farrell+09}
Farrell, S.~A., Webb, N.~A., Barret, D., Godet, O., \& Rodrigues, J.~M. 2009,
  Nat., 460, 73

\bibitem[{Feng \& Kaaret(2006)}]{Feng+06}
Feng, H., \& Kaaret, P. 2006, ApJ, 650, L75

\bibitem[{Feng \& Soria(2011)}]{FengSoria}
Feng, H., \& Soria, R. 2011, New Astronomy Reviews, 55, 166

\bibitem[{Gladstone {et~al.}(2009)Gladstone, Roberts, \& Done}]{Gladstone+09}
Gladstone, J.~C., Roberts, T.~P., \& Done, C. 2009, MNRAS, 397, 1836

\bibitem[{Gladstone {et~al.}(2011)Gladstone, Roberts, \& Done}]{Gladstone+11}
---. 2011, Astron. Nachr., 332, 345

\bibitem[{Godet {et~al.}(2012)Godet, Plazolles, Kawaguchi, Lasota, Barret,
  Farrell, Braito, Servillat, Webb, \& Gehrels}]{Godet+12}
Godet, O., Plazolles, B., Kawaguchi, T., {et~al.} 2012, ApJ, 752, 34

\bibitem[{Gon{\c c}alves \& Soria(2006)}]{GoncalvesSoria06}
Gon{\c c}alves, A.~C., \& Soria, R. 2006, MNRAS, 371, 673

\bibitem[{Harrison {et~al.}(2013)Harrison, Craig, Christensen, Hailey, Zhang,
  Boggs, Stern, Cook, Forster, Giommi, Grefenstette, Kim, Kitaguchi, Koglin,
  Madsen, Mao, Miyasaka, Mori, Perri, Pivovaroff, Puccetti, Rana, Westergaard,
  Willis, Zoglauer, An, Bachetti, Barri{\`e}re, Bellm, Bhalerao, Brejnholt,
  Fuerst, Liebe, Markwardt, Nynka, Vogel, Walton, Wik, Alexander, Cominsky,
  Hornschemeier, Hornstrup, Kaspi, Madejski, Matt, Molendi, Smith, Tomsick,
  Ajello, Ballantyne, Balokovi{\'c}, Barret, Bauer, Blandford, Brandt,
  Brenneman, Chiang, Chakrabarty, Chenevez, Comastri, Dufour, Elvis, Fabian,
  Farrah, Fryer, Gotthelf, Grindlay, Helfand, Krivonos, Meier, Miller,
  Natalucci, Ogle, Ofek, Ptak, Reynolds, Rigby, Tagliaferri, Thorsett,
  Treister, \& Urry}]{nustar13}
Harrison, F.~A., Craig, W.~W., Christensen, F.~E., {et~al.} 2013, ApJ, 770, 103

\bibitem[{Heil {et~al.}(2009)Heil, Vaughan, \& Roberts}]{Heil+09}
Heil, L.~M., Vaughan, S., \& Roberts, T.~P. 2009, MNRAS, 397, 1061

\bibitem[{Houck \& Denicola(2000)}]{Houck+00}
Houck, J.~C., \& Denicola, L.~A. 2000, ADASS, 216, 591

\bibitem[{Jansen {et~al.}(2001)Jansen, Lumb, Altieri, Clavel, Ehle, Erd,
  Gabriel, Guainazzi, Gondoin, Much, Munoz, Santos, Schartel, Texier, \&
  Vacanti}]{xmm01}
Jansen, F., Lumb, D., Altieri, B., {et~al.} 2001, A{\&}A, 365, L1

\bibitem[{Joye \& Mandel(2003)}]{ds9}
Joye, W.~A., \& Mandel, E. 2003, ADASS, 295, 489

\bibitem[{Kajava \& Poutanen(2009)}]{KajavaPoutanen09}
Kajava, J. J.~E., \& Poutanen, J. 2009, MNRAS, 398, 1450

\bibitem[{Kalberla {et~al.}(2005)Kalberla, Burton, Hartmann, Arnal, Bajaja,
  Morras, \& P{\"o}ppel}]{Kalberla+05}
Kalberla, P. M.~W., Burton, W.~B., Hartmann, D., {et~al.} 2005, A{\&}A, 440,
  775

\bibitem[{Kawaguchi(2003)}]{Kawaguchi03}
Kawaguchi, T. 2003, ApJ, 593, 69

\bibitem[{King(2004)}]{King+04}
King, A. 2004, Nuclear Physics B Proceedings Supplements, 132, 376

\bibitem[{Kubota {et~al.}(2005)Kubota, Ebisawa, Makishima, \&
  Nakazawa}]{Kubota+05}
Kubota, A., Ebisawa, K., Makishima, K., \& Nakazawa, K. 2005, ApJ, 631, 1062

\bibitem[{Laor(1991)}]{Laor91}
Laor, A. 1991, ApJ, 376, 90

\bibitem[{Leahy {et~al.}(1983)Leahy, Darbro, Elsner, Weisskopf, Kahn,
  Sutherland, \& Grindlay}]{Leahy+83}
Leahy, D.~A., Darbro, W., Elsner, R.~F., {et~al.} 1983, ApJ, 266, 160

\bibitem[{McHardy(2010)}]{McHardy10}
McHardy, I. 2010, The Jet Paradigm, 794, 203

\bibitem[{M{\'e}ndez {et~al.}(2002)M{\'e}ndez, Davis, Moustakas, Newman,
  Madore, \& Freedman}]{Mendez+02}
M{\'e}ndez, B., Davis, M., Moustakas, J., {et~al.} 2002, The Astronomical
  Journal, 124, 213

\bibitem[{Mewe \& Gronenschild(1981)}]{MeweGronenshild81}
Mewe, R., \& Gronenschild, E. H. B.~M. 1981, A{\&}A Supp. Ser., 45, 11

\bibitem[{Middleton {et~al.}(2011{\natexlab{a}})Middleton, Roberts, Done, \&
  Jackson}]{Middleton+11}
Middleton, M.~J., Roberts, T.~P., Done, C., \& Jackson, F.~E.
  2011{\natexlab{a}}, MNRAS, 411, 644

\bibitem[{Middleton {et~al.}(2011{\natexlab{b}})Middleton, Sutton, \&
  Roberts}]{Middleton+11M33}
Middleton, M.~J., Sutton, A.~D., \& Roberts, T.~P. 2011{\natexlab{b}}, MNRAS,
  417, 464

\bibitem[{Middleton {et~al.}(2012)Middleton, Sutton, Roberts, Jackson, \&
  Done}]{Middleton+12}
Middleton, M.~J., Sutton, A.~D., Roberts, T.~P., Jackson, F.~E., \& Done, C.
  2012, MNRAS, 420, 2969

\bibitem[{Middleton {et~al.}(2013)Middleton, Miller-Jones, Markoff, Fender,
  Henze, Hurley-Walker, Scaife, Roberts, Walton, Carpenter, Macquart, Bower,
  Gurwell, Pietsch, Haberl, Harris, Daniel, Miah, Done, Morgan, Dickinson,
  Charles, Burwitz, Della~Valle, Freyberg, Greiner, Hernanz, Hartmann,
  Hatzidimitriou, Riffeser, Sala, Seitz, Reig, Rau, Orio, Titterington, \&
  Grainge}]{Middleton+13}
Middleton, M.~J., Miller-Jones, J. C.~A., Markoff, S., {et~al.} 2013, Nat.,
  493, 187

\bibitem[{Miller {et~al.}(2003)Miller, Fabbiano, Miller, \& Fabian}]{Miller+03}
Miller, J.~M., Fabbiano, G., Miller, M.~C., \& Fabian, A.~C. 2003, ApJ, 585,
  L37

\bibitem[{Miller {et~al.}(2004)Miller, Fabian, \& Miller}]{Miller+04}
Miller, J.~M., Fabian, A.~C., \& Miller, M.~C. 2004, ApJ, 614, L117

\bibitem[{Miller {et~al.}(2013)Miller, Walton, King, Reynolds, Fabian, Miller,
  \& Reis}]{Miller+13}
Miller, J.~M., Walton, D.~J., King, A.~L., {et~al.} 2013, ApJL, 776, L36

\bibitem[{Mineshige {et~al.}(1994)Mineshige, Hirano, Kitamoto, Yamada, \&
  Fukue}]{Mineshige+94}
Mineshige, S., Hirano, A., Kitamoto, S., Yamada, T.~T., \& Fukue, J. 1994, ApJ,
  426, 308

\bibitem[{Mitsuda {et~al.}(1984)Mitsuda, Inoue, Koyama, Makishima, Matsuoka,
  Ogawara, Suzuki, Tanaka, Shibazaki, \& Hirano}]{Mitsuda+84}
Mitsuda, K., Inoue, H., Koyama, K., {et~al.} 1984, Astronomical Society of
  Japan, 36, 741

\bibitem[{Mitsuda {et~al.}(2007)Mitsuda, Bautz, Inoue, Kelley, Koyama, Kunieda,
  Makishima, Ogawara, Petre, Takahashi, Tsunemi, White, Anabuki, Angelini,
  Arnaud, Awaki, Bamba, Boyce, Brown, Chan, Cottam, Dotani, Doty, Ebisawa,
  Ezoe, Fabian, Figueroa, Fujimoto, Fukazawa, Furusho, Furuzawa, Gendreau,
  Griffiths, Haba, Hamaguchi, Harrus, Hasinger, Hatsukade, Hayashida, Henry,
  Hiraga, Holt, Hornschemeier, Hughes, Hwang, Ishida, Ishisaki, Isobe, Itoh,
  Iyomoto, Kahn, Kamae, Katagiri, Kataoka, Katayama, Kawai, Kilbourne,
  Kinugasa, Kissel, Kitamoto, Kohama, Kohmura, Kokubun, Kotani, Kotoku, Kubota,
  Madejski, Maeda, Makino, Markowitz, Matsumoto, Matsumoto, Matsuoka,
  Matsushita, McCammon, Mihara, Misaki, Miyata, Mizuno, Mori, Mori, Morii,
  Moseley, Mukai, Murakami, Murakami, Mushotzky, Nagase, Namiki, Negoro,
  Nakazawa, Nousek, Okajima, Ogasaka, Ohashi, Oshima, Ota, Ozaki, Ozawa,
  Parmar, Pence, Porter, Reeves, Ricker, Sakurai, Sanders, Senda, Serlemitsos,
  Shibata, Soong, Smith, Suzuki, Szymkowiak, Takahashi, Tamagawa, Tamura,
  Tamura, Tanaka, Tashiro, Tawara, Terada, Terashima, Tomida, Torii, Tsuboi,
  Tsujimoto, Tsuru, Turner, Ueda, Ueno, Ueno, Uno, Urata, Watanabe, Yamamoto,
  Yamaoka, Yamasaki, Yamashita, Yamauchi, Yamauchi, Yaqoob, Yonetoku, \&
  Yoshida}]{suzaku07}
Mitsuda, K., Bautz, M., Inoue, H., {et~al.} 2007, PASJ, 59, 1

\bibitem[{Mizuno {et~al.}(2007)Mizuno, Miyawaki, Ebisawa, Kubota, Miyamoto,
  Winter, Ueda, Isobe, Dewangan, Done, Griffiths, Haba, Kokubun, Kotoku,
  Makishima, Matsushita, Mushotzky, Namiki, Petre, Takahashi, Tamagawa, \&
  Terashima}]{Mizuno+07}
Mizuno, T., Miyawaki, R., Ebisawa, K., {et~al.} 2007, PASJ, 59, 257

\bibitem[{Nowak(2005)}]{Nowakrant}
Nowak, M. 2005, Ap{\&}SS, 300, 159

\bibitem[{Pakull \& Mirioni(2002)}]{PakullMirioni02}
Pakull, M. W., \& Mirioni, L. 2002, in the proceedings of the symposium ``New Visions of the X-ray Universe in the XMM-Newton and Chandra Era'', 26-30 November 2001, ESTEC, The Netherlands

\bibitem[{Pintore \& Zampieri(2011)}]{Pintore+11}
Pintore, F., \& Zampieri, L. 2011, Astron. Nachr., 332, 337

\bibitem[{Pintore \& Zampieri(2012)}]{Pintore+12}
---. 2012, MNRAS, 420, 1107

\bibitem[{Poutanen {et~al.}(2007)Poutanen, Lipunova, Fabrika, Butkevich, \&
  Abolmasov}]{Poutanen+07}
Poutanen, J., Lipunova, G., Fabrika, S., Butkevich, A.~G., \& Abolmasov, P.
  2007, MNRAS, 377, 1187

\bibitem[{Ramsey {et~al.}(2006)Ramsey, Williams, Gruendl, Chen, Chu, \&
  Wang}]{Ramsey+06}
Ramsey, C.~J., Williams, R.~M., Gruendl, R.~A., {et~al.} 2006, ApJ, 641, 241

\bibitem[{Remillard \& McClintock(2006)}]{Remillard+06}
Remillard, R.~A., \& McClintock, J.~E. 2006, ARA{\&}A, 44, 49

\bibitem[{Roberts(2007)}]{Roberts07}
Roberts, T.~P. 2007, Ap{\&}SS, 311, 203

\bibitem[{Ross \& Fabian(2005)}]{reflionx05}
Ross, R.~R., \& Fabian, A.~C. 2005, MNRAS, 358, 211

\bibitem[{Shakura \& Sunyaev(1973)}]{SS73}
Shakura, N.~I., \& Sunyaev, R.~A. 1973, A{\&}A, 24, 337

\bibitem[{Soria {et~al.}(2004)Soria, Motch, Read, \& Stevens}]{Soria+04}
Soria, R., Motch, C., Read, A.~M., \& Stevens, I.~R. 2004, A{\&}A, 423, 955

\bibitem[{Stobbart {et~al.}(2006)Stobbart, Roberts, \& Wilms}]{Stobbart+06}
Stobbart, A.-M., Roberts, T.~P., \& Wilms, J. 2006, MNRAS, 368, 397

\bibitem[{Straub {et~al.}(2013)Straub, Done, \& Middleton}]{Straub+13}
Straub, O., Done, C., \& Middleton, M. 2013, A{\&}A, 553, 61

\bibitem[{Sutton {et~al.}(2013)Sutton, Roberts, \& Middleton}]{Sutton+13}
Sutton, A.~D., Roberts, T.~P., \& Middleton, M.~J. 2013, MNRAS, 435, 1758

\bibitem[{Swartz {et~al.}(2004)Swartz, Ghosh, Tennant, \& Wu}]{Swartz+04}
Swartz, D.~A., Ghosh, K.~K., Tennant, A.~F., \& Wu, K. 2004, ApJ Supp. Ser.,
  154, 519

\bibitem[{Titarchuk(1994)}]{comptt94}
Titarchuk, L. 1994, ApJ, 434, 570

\bibitem[{Uttley \& McHardy(2001)}]{UttleyMcHardy01}
Uttley, P., \& McHardy, I.~M. 2001, MNRAS, 323, L26

\bibitem[{van~der Klis(1989)}]{VDK89}
van~der Klis, M. 1989, in Timing Neutron Stars: proceedings of the NATO
  Advanced Study Institute on Timing Neutron Stars held April 4-15, 27

\bibitem[{Vaughan {et~al.}(2003)Vaughan, Edelson, Warwick, \&
  Uttley}]{Vaughan+03}
Vaughan, S., Edelson, R., Warwick, R.~S., \& Uttley, P. 2003, MNRAS, 345, 1271

\bibitem[{Verner {et~al.}(1996)Verner, Ferland, Korista, \&
  Yakovlev}]{Verner+96}
Verner, D.~A., Ferland, G.~J., Korista, K.~T., \& Yakovlev, D.~G. 1996,
  Astrophysical Journal v.465, 465, 487

\bibitem[{Vierdayanti {et~al.}(2006)Vierdayanti, Mineshige, Ebisawa, \&
  Kawaguchi}]{Vierdayanti+06}
Vierdayanti, K., Mineshige, S., Ebisawa, K., \& Kawaguchi, T. 2006, PASJ, 58,
  915

\bibitem[{Walton {et~al.}(2011{\natexlab{a}})Walton, Gladstone, Roberts,
  Fabian, Caballero-Garcia, Done, \& Middleton}]{Walton+11}
Walton, D.~J., Gladstone, J.~C., Roberts, T.~P., {et~al.} 2011{\natexlab{a}},
  MNRAS, 414, 1011

\bibitem[{Walton {et~al.}(2012)Walton, Miller, Reis, \& Fabian}]{Walton+12}
Walton, D.~J., Miller, J.~M., Reis, R.~C., \& Fabian, A.~C. 2012, MNRAS, 426,
  473

\bibitem[{Walton {et~al.}(2011{\natexlab{b}})Walton, Roberts, Mateos, \&
  Heard}]{Walton+11cat}
Walton, D.~J., Roberts, T.~P., Mateos, S., \& Heard, V. 2011{\natexlab{b}},
  MNRAS, 416, 1844

\bibitem[{Watarai \& Fukue(1999)}]{WataraiFukue99}
Watarai, K.-y., \& Fukue, J. 1999, PASJ, 51, 725

\bibitem[{Weisskopf {et~al.}(2002)Weisskopf, Brinkman, Canizares, Garmire,
  Murray, \& {van Speybroeck, L. P.}}]{chandra02}
Weisskopf, M.~C., Brinkman, B., Canizares, C., {et~al.} 2002, The Publications
  of the Astronomical Society of the Pacific, 114, 1

\bibitem[{Wilms {et~al.}(2000)Wilms, Allen, \& McCray}]{Wilms+00}
Wilms, J., Allen, A., \& McCray, R. 2000, ApJ, 542, 914

\end{thebibliography}

\end{document}